\documentclass[useAMS,usedcolumn]{mn2e}
\usepackage{rotating}
\usepackage{lscape}
\usepackage{latexsym}
\usepackage{graphicx}
\usepackage{amssymb}
\usepackage{lscape}
\usepackage{pifont}
\usepackage{amsfonts}
\usepackage{amsmath}
\usepackage{float}
\usepackage{times}
\usepackage{multirow}
\usepackage{flafter}
\usepackage{placeins}
\usepackage{afterpage}
\usepackage{preview}
\usepackage{sidecap}
\usepackage{setspace}
\usepackage{subfigure}
\usepackage{textcomp}
\usepackage{bbding}
\usepackage{epsf}

\newcommand{\NII}{[N\,{\sc ii}]}

\newcommand{\SII}{[S\,{\sc ii}]}

\newcommand{\OIII }{[O\,{\sc iii}]}

\newcommand{\OII}{[O\,{\sc ii}]}

\def\p0{\phantom{0}}

\def\cm3{cm$^{-3}$}

\def\12{$^{12}$CO}
\def\13co{$^{13}$CO}

\def\la{{\hbox{$\lesssim$}}}

\begin{document}
\title[The nearest collisional ring galaxy]{Kathryn's Wheel: A spectacular galaxy collision discovered in the Galactic
neighbourhood\thanks{Named after the wife of the 2nd author}}
\author[Parker et al.]{Quentin A. Parker$^{1,2}$\thanks{E-mail:
quentinp@hku.hk}, Albert A. Zijlstra$^{3}$\thanks{E-mail:
a.zijlstra@manchester.ac.uk}, 
Milorad Stupar$^4$,
Michelle Cluver$^5$,\newauthor  David J. Frew$^1$, George Bendo$^{3}$ \& Ivan Boji\v{c}i\'c$^1$ \\
$^{1}$ Department of Physics, Chong Yuet Ming Physics building, University of Hong Kong, Hong Kong\\ 
$^{2}$ Australian Astronomical Observatory, PO Box 915, North Ryde NSW 1670, Australia\\
$^{3}$ Jodrell Bank centre for Astrophysics, School of Physics \&\ Astronomy, University of Manchester, Oxford Road, Manchester M13 9PL, UK\\
$^{4}$ University of Western Sydney, Locked Bag 1797, Penrith South, DC, NSW 1797, Australia\\
$^{5}$ University of the Western Cape, Robert Sobukwe Road, Bellville, 7535, South Africa
}

\date{Accepted . Received 2015}

\maketitle

\begin{abstract}
We report the discovery of the closest collisional ring galaxy to the Milky Way.  Such rare systems occur due to ``bulls-eye'' encounters between two reasonably 
matched galaxies. The recessional velocity of about 840\,km\,s$^{-1}$ is low enough that it was detected in the AAO/UKST Survey for Galactic H$\alpha$ emission.  
The distance is only 10.0\,Mpc and the main galaxy shows a full ring of star forming knots, 6.1\,kpc in 
diameter surrounding a quiescent disk. The smaller assumed ``bullet" galaxy also shows vigorous star formation.  
The spectacular nature of the object had been overlooked because of its location in the Galactic plane and proximity to a bright star and even though it is the
60$^{\rm  th}$ brightest galaxy in the HI Parkes All Sky Survey (HIPASS) H\,{\sc i} survey. 

The overall system has a physical size of $\sim$15\,kpc, a total mass of $M_\ast = 6.6\times 10^9$\,M$_\odot$ (stars + H{\sc i}), a metallicity of [O/H]$\sim-0.4$, and 
a star formation rate of 0.2-0.5\,M$_\odot$\,yr$^{-1}$, making it a Magellanic-type system.  Collisional ring galaxies therefore extend to much lower galaxy 
masses than commonly assumed. We derive a space density for such systems of $7 \times 10^{-5}\,\rm Mpc^{-3}$, an order of magnitude higher than previously 
estimated. This suggests Kathryn's Wheel is the nearest such system.  We present discovery images, CTIO 4-m telescope narrow-band follow-up images and 
spectroscopy for selected emission components.  Given its proximity and modest extinction along the line of sight, this spectacular system provides an ideal target for 
future high spatial resolution studies of such systems and for direct detection of its stellar populations.
\end{abstract}


\section{Introduction}
Among the structures shown by interacting galaxies (Barnes \& Hernquist 1992), the so-called collisional ring galaxies (Theys \& Spiegel 1976, 1977) are a rare 
and most distinctive sub-class  (Appelton \& Struck-Marcell 1996).  Collisional ring galaxies occur from an almost head-on collision between a 
smaller galaxy travelling along the minor axis of a larger galaxy passing through, or close to, its centre (Appleton \& Struck-Marcell 1996).  The shock wave sweeps up and expels gas from the system, usually leaving a gas-poor galaxy behind (Freeman \& De Vaucoleurs 1974). 

The pre-eminent example is the Cartwheel galaxy (ESO\,350-40; AM\,0035-335) at a distance of 124\,Mpc.  The Cartwheel was discovered by Zwicky (1941), and 
studied in-depth by Fosbury \& Hawarden (1977) and Amram et al. (1998).  The Cartwheel shows a spectacular ring of star formation, with interior spokes 
surrounding  the host galaxy.  Other well-known collisional systems include Arp\,148 (Mayall's Object) at 148\,Mpc  (Smith 1941; Burbidge, Burbidge \& Prendrgast 
1964; Arp 1966), IC~298 (Arp\,147; Gerber, Lamb \& Balsara 1992) at 135\,Mpc, and AM\,0644-741 (ESO\,34-11 or the ``Lindsay-Shapley Ring'') at 87\,Mpc  (Lindsay \& Shapley 
1960; Graham  1974; Arp \&  Madore 1987).   This latter object is often considered to be a typical exemplar of 
evolved ringed galaxies (e.g. Higdon, Higdon \& Rand 2011).

While star-forming events resulting from galactic collisions may be more common than usually thought (e.g. Block et al. 2006), spectacular cartwheel-type galaxies, 
showing complete rings (but not necessarily with evident spokes), appear to be very rare in volume-limited surveys.  The previous nearest example was the Vela Ring 
galaxy (AM\,1006-380 = ESO\,316-43), at 72\,Mpc (Dennefeld, Lausten \& Materne 1979).   Partial ring galaxies are more numerous.   Probably the nearest in the 
Madore, Nelson \& Petrillo (2009) catalogue is AM\,0322-374 at 20\,Mpc.  UGC\,9893, at 11\,Mpc (Kennicutt et al. 2008), is also listed as a collisional system, but it lacks 
a clear ring structure (Gil de Paz \& Madore 2003), appearing indistinct in available imagery.  Collisional ring galaxy systems can be identified in high-
resolution, deep H$\alpha$ imagery where the H\,{\sc ii} regions of the ring are readily visible.  However, H$\alpha$ galaxy surveys such as Gil de Paz \& Madore (2003), 
James et al. (2004), and Kennicutt et al. (2008) have not added to the modest number of cartwheel-type galaxies currently known and there are $<$20 
convincing systems, e.g. Marston \& Appleton 1995. Bosch et al. (2015) present the most recent detailed work on such systems for an apparently empty ring galaxy system ESO\,474\,G040
at $\sim78$\,Mpc, thought to have arisen from a recent merger of two disk galaxies.

Collisional rings systems offer an excellent environment to study the evolution of density-wave induced starbursts (Marston \& Appleton 1995; Romano, Mayya \& 
Vorobyov 2008), as well as the dynamic interactions between the individual galaxies (e.g. Smith et al. 2012).
A closer example of a collisional ring system would allow its star formation, stellar distribution and evolution to be studied in detail, spatially, kinematically and 
chromatically.   Here we present the discovery of a new collisional ring galaxy at a distance of only 10\,Mpc.  Our paper is organised as follows. In section~2 we present 
some background and describe the discovery of the ring around ESO\,179-13. In section~3 we present our follow-up high-resolution imaging and other archival multi-
wavelength images. Optical spectroscopy is presented in section~4. Parameters for the system, such as masses and star formation rates, are provided in section~5. 
Section~6 presents our discussion and give our summary in Section~7.

\section{Background and Discovery}

\subsection{Background} 
ESO\,179-13 is noted as an interacting double galaxy system in SIMBAD, appearing to consist of a near edge-on late-type spiral with another more chaotic system 
$\sim$\,80\,arcsec to the north-east.  It was first identified by Sersic (1974) and independently re-discovered by Lauberts (1982), Longmore et al. (1982) and Corwin, de 
Vaucouleurs \& de Vaucouleurs (1985).  A detailed $B$-band image was published by Laustsen, Madsen \& West (1987). 

Woudt \& Kraan-Korteweg (2001) identify the most prominent galaxy as WKK\,7460 (component\,A hereafter), and WKK\,7463 for the secondary (component B). 
A small, third galaxy is also identified as WKK\,7457 to the west of component\,A (component\,C). It agrees with our own positional 
determination to 1\,arcsec.
The system, centred on component\,A, is entry 391 in the H$\alpha$ catalogue of Kennicutt et al. (2008), but they do not list integrated H$\alpha$ (+\,[N\,II]) flux or 
luminosity estimates.  Lee et al. (2011), in a related paper, included this system in their Galaxy Evolution Explorer (GALEX) ultraviolet imaging survey of local volume galaxies with a quoted $B_T$ = 
15, but no near-UV or far-UV fluxes are quoted. 
Owing to its proximity to the Galactic plane, the galaxy was not observed by GALEX prior to mission completion (Bianchi 2014).

Component\,A has been classified as an SB\,(s)\,m spiral in the RC3 catalogue (de Vaucouleurs et al. 1991) and similarly classified as SB\,(s)\,dm 
system (T = 7.5) by Buta (1995).  These independent classifications treat the system as a distinct, late-type spiral galaxy, albeit in an interaction.  It can be best 
described as a Magellanic dwarf or a Magellanic spiral, similar to the LMC.

A range of heliocentric radial velocities for component\,A are available in the literature from both optical spectroscopic and H{\sc i} measurements.  The most recent 
optical results are Strauss et al. (1992) who quote 775$\pm$36 km\,s$^{-1}$, Di Nella et al. (1997), 750$\pm$70 km\,s$^{-1}$, and Woudt et al. (1999), 757$\pm$50 km
\,s$^{-1}$.   These are in good agreement so we take an unweighted average of 761\,km\,s$^{-1}$ ($\sigma$=11\,km\,s$^{-1}$) for component\,A.   
 
The earliest H{\sc i} heliocentric velocities  towards the system are 836$\pm$5\,km\,s$^{-1}$ by Longmore et al. (1982),  840$\pm$20\,km\,s$^{-1}$  by Tully 
(1988) and 843$\pm$9\,km\,s$^{-1}$  by de Vaucouleurs et al. (1991).  The Parkes H{\sc i} `HIPASS' Multibeam receiver programme gives 843$\pm$5\,km\,s
$^{-1}$  (HIPASS catalogue; Meyer et al. 2004) where it is noted as the 60$^{\rm th}$ brightest in terms of integrated flux density (Koribalski et al. 
2004). A deeper, pointed Parkes observation was obtained by Schr\"oder et al. (2009).  They give parameters for component\,A, but note that a contribution from 
component\,B is likely.  Their heliocentric H{\sc i} radial velocity is 842\,km\,s$^{-1}$ from the H{\sc i} profile mid-point, with a velocity width of 222\,km\,s$^{-1}$ at 20\% 
of peak. The integrated H{\sc i} flux density is 113.8\,$\pm$\,3.7\,Jy\,km\,s$^{-1}$, in good agreement with the HIPASS value of 100.  All the H{\sc i} velocities are 
consistent within small errors.  These values include the combined contribution from all the gas in the full system.  We adopt 842$\pm$5\,km\,s$^{-1}$ 
for the combined H{\sc i} system velocity.

Kennicutt et al. (2008) give a distance of 9\,Mpc based on a Virgocentric flow model distance. We adopt a slightly further distance of 10.0\,$\pm 1.0$ Mpc, based on 
Kennicutt et al. (2008), but scaling from a Hubble constant H$_0$ of 75 to 68 km\,s$^{-1}$\,Mpc$^{-1}$ due to the latest WMAP (Hinshaw et al. 2013) and Planck results 
(Ade  et al. 2014).  At this distance, 1\,arcsec equates to 48.5\,pc.   Wouldt \& Kraan-Korteweg (2001) give a modest colour excess of $E$(B-V)$\,=\,0.25$~mag in the surrounding 
field, giving $A_V=0.78$~mag while  Schlafly \& Finkbeiner (2011) give $A_V=0.75$~mag for Galactic extinction in this general direction, which we adopt hereafter.  We will 
also use $A_{H\alpha} = 0.61$~mag based on the reddening law from Howarth (1983). 

Apart from the basic information above, the system has been little studied because of its location in a crowded, low-latitude area near the Galactic 
plane ($|b| =$-8 degrees) and proximity to the bright, 7.7~mag A0\,IV star, HD\,150915, located 72\,arcsec due south.  There are actually two bright stars here, but only 
HD\,150915 is listed in SIMBAD. The second star is $\sim$10\,arcsec north of HD\,150915 and $\sim$2\,magnitudes fainter.  It was undoubtedly blended with HD\,
150915 in the original photographic imagery.

\subsection{Discovery of the Ring}
\begin{figure*}
\includegraphics[width=17.5cm,height=6cm]{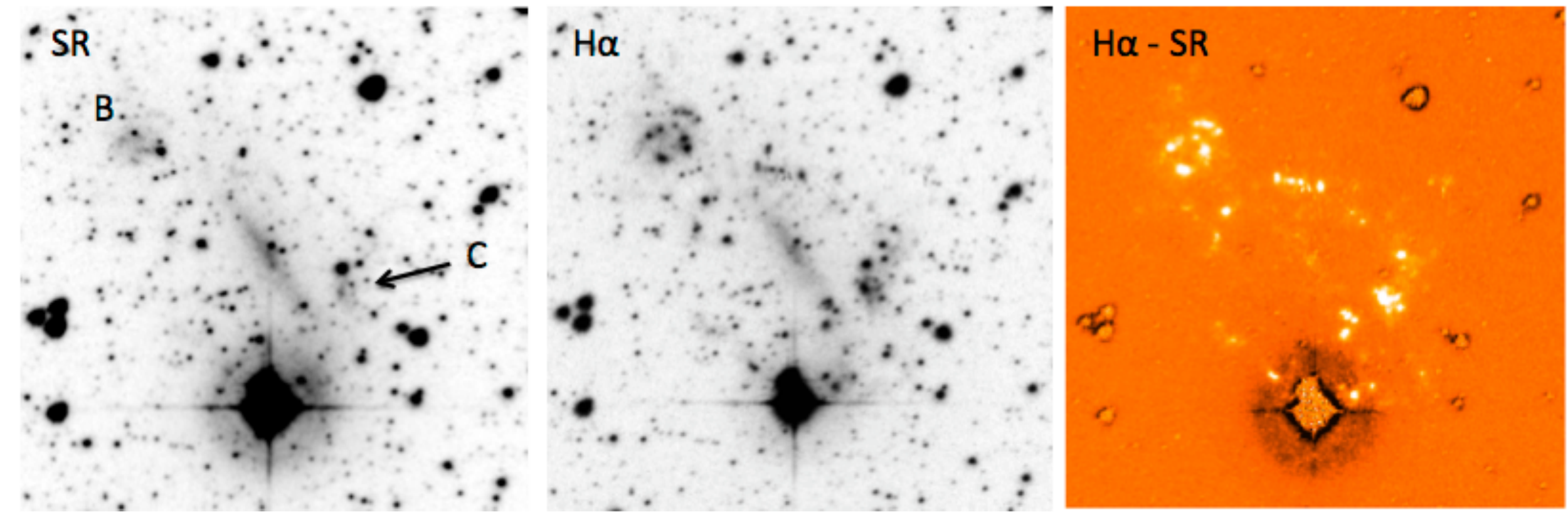}
\caption{Left panel: The $4\times4$~arcminute SHS  broad-band short-red `SR' image centred on component\,A at  16h47m19.7s, -57$^{o}$26'31'' (J2000).  
North is top and east left. Mid panel: The matching H$\alpha$ image.
Right panel: The continuum subtracted image obtained by taking the SR from the SHS H$\alpha$ (+\,[N\,II]) image scaled at 99.5\%. 
North is at the top and east to the left. Component\,B is labelled to the north-east and  component\,C is indicated 
with an arrow.}
\label{SHS}
\end{figure*}

The low heliocentric velocity meant the system was discovered as a strong H$\alpha$ emitter in the Anglo-Australian Observatory (AAO) / U.K. Schmidt Telescope (UKST)  
SuperCOSMOS H$\alpha$ survey of the Southern Galactic Plane (SHS; Parker et al. 2005) during searches for Galactic planetary nebulae (Parker et al. 2006).  
The bandpass of the survey filter (Parker \& Bland-Hawthorn 1998) 
includes all velocities of Galactic gas and emission from extragalactic objects out to the Virgo and Fornax galaxy clusters at $\sim$1200\,km\,s$^{-1}$.  

Fig. \ref{SHS} shows the 4~$\times$~4\,arcmin discovery images from the SHS survey data centred on the system. From left to right there is the H$\alpha$ image, the 
matching broadband  short-red `SR'  image and the image obtained by taking the SR image from the H$\alpha$ image (effectively a continuum subtracted image). 
Component\,A is at the centre. Component\,B is labelled to the north-east while the smaller compact component\,C is indicated with an arrow.
 
The SHS continuum subtracted image reveals a spectacular ring of emission knots not obvious, without context, even in the H$\alpha$ image 
while the main component\,A galaxy has almost disappeared. The ring is approximately (but not perfectly) centred on component\,A. Surprisingly, given its previous 
spiral morphological classification, component\,A does not show up well in the SHS continuum subtracted image and appears to show little star formation. This  indicates 
that along with its apparently highly elongated nature (so not an elliptical) it is has been largely stripped of gas. The smaller, more irregular galaxy (component\,B)  
shows extensive, clumpy H$\alpha$ emission.  A third, faint compact system (component\,C) is seen 38\,arcsec to the west of component\,A, 
identified as WKK\,7457. This object is clear in the broad-band $B$, $R$ and $I$-band SuperCOSMOS images and is also undergoing star-formation.

The strong H{\sc i} detection noted earlier indicates that a significant, neutral H{\sc i} component still remains within the overall system environment.   The lack of much 
star formation in component\,A suggests that most of this H{\sc i} may be located outside of its inner disk.

\section{Follow-up imaging and multi-wavelength comparisons}

Discovery of this collisional ring system led us first to investigate and compile existing multi-wavelength data from the archives.  We then obtained deeper, higher resolution 
imaging and  spectroscopy of the major emission components (see below), to provide estimates of some key physical parameters for this system.  

\subsection{CTIO 4m-telescope Imaging}
High resolution imagery at the Blanco 4-m telescope at CTIO was obtained in June 2008 using the wide field mosaic camera.  We used five filters: $V$-band, H$\alpha
$, H$\alpha$ off-band (80\AA\ redward of H$\alpha$), \OIII\,  , and a (wider) \OIII\,   off-band filter centred at 5300\AA.   The field of view is 30\,arcmin on a side and the 
plate scale 0.26\,arcsec\,px$^{-1}$.  The seeing measured from the data frames was typically 1.2\,arcsec.  Exposures were 5 minutes for the emission-line and off-band 
filters, and 60\,s for the $V$-band filter.  The airmass during the exposures averaged 1.3.  To subtract the continuum, the H$\alpha$ off-band image was scaled down by 
a factor of 0.88, and the \OIII\,   off-band by a factor of 0.14, to compensate for the filter widths.  These factors were established by minimizing residuals for a
set of selected field stars.  For \OIII\,   that works very well and results are in good agreement with the filter curve.  For H$\alpha$ this can be an issue if the chosen stars 
have H$\alpha$ in absorption. This does not appear to be the case for the field stars chosen which provide a consistent scale-factor to $\sim2\%$. It is best to minimize stellar 
residuals rather than filter curves as stars are a more important contribution to the continuum than bound-free emission from H{\sc i} regions.

\begin{figure*}
\includegraphics[width=17.5cm,height=6cm]{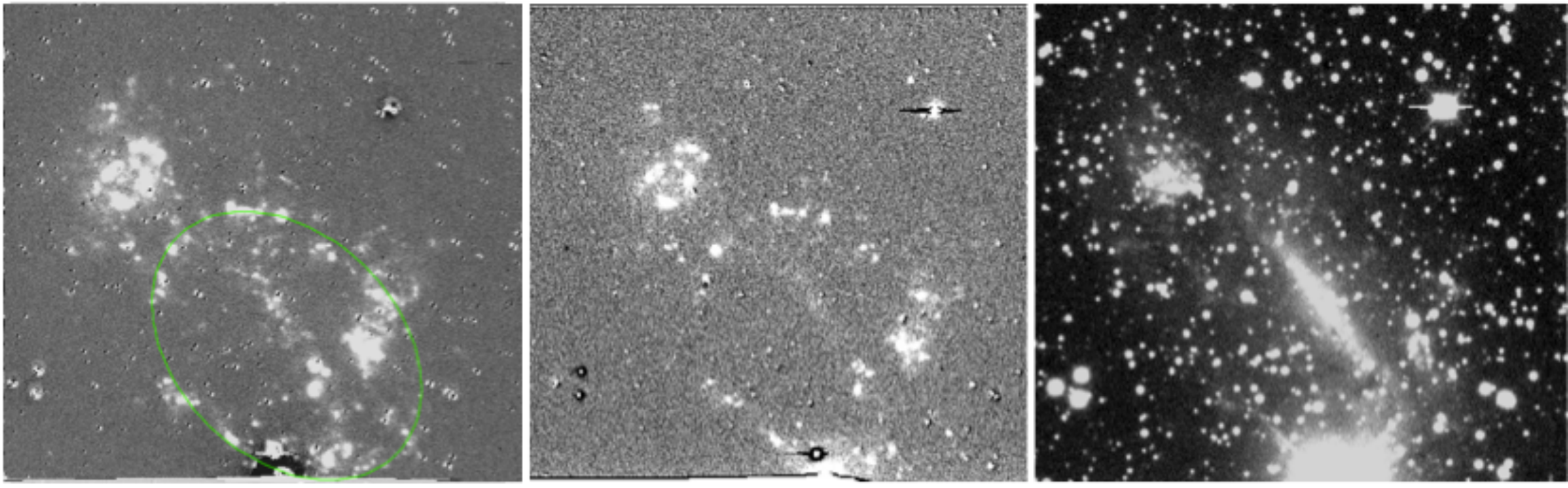}
\caption{Selected $3\times3$\,arcmin regions from the CTIO 4m imaging centred at 16h47m21s, -57$^{o}$26'03'' (J2000). Left panel: the CTIO continuum subtracted
H$\alpha$. The green oval has a major axis diameter of 127\,arcsec ($\sim$6.2\,kpc) and a minor axis of 91\,arcsec ($\sim$4.4\,kpc). 
Mid panel: matching CTIO continuum subtracted \OIII\,   image. Both these images are at 90\% scaling to better reveal the emission otherwise the 7.7th magnitude 
star dominates. Right panel: CTIO \OIII\,   off-band image. This represents the $V$-band starlight without any \OIII\,   and H$\beta$ emission line contribution. 
Dust lanes can be clearly seen at the southern tip of component\,A. The more diffusely distributed starlight can also be seen in and 
around components\,A and B.}
\label{CTIO}
\end{figure*}

A montage of these CTIO images is shown in Fig. \ref{CTIO}. They reveal the distribution and variety of the 
emission structures as well as data for quantitative flux estimates.
From these deeper, higher resolution CTIO H$\alpha$ and \OIII\,   continuum subtracted images the distribution of ionised gas is seen to be far more complex than is 
evident from the SHS discovery data. 

The leftmost panel is a $3\times3$\,arcminute CTIO H$\alpha$ image with the off-band frame subtracted.  We avoid the bright star to the south where CCD blooming is 
serious. Emission associated with component\,C and at the southern tip of component\,A's disk previously gave 
the impression of a circular ring in the SHS as the bright star obscured emission further south. The higher resolution CTIO data reveals an elliptical emission ring with a 
major-axis diameter of 127\,arcsec ($\sim$6.2\,kpc). The inscribed oval was positioned to fit the prominent emission
now seen to extend further south.  The ellipse is not centred on component\,A, but at 16h47m19.5s, $-$57$^{o}$26' 44'' (J2000), 
$\sim$13\,arcsec (or $\sim$0.6\,kpc) south-south-east. 
Some low-level H$\alpha$ emission is now seen across component\,A, surrounding component\,B and with faint, localised emission features and blobs
around the entire system. An interesting feature is the elongated nature of some of these emission knots in a north-east direction from A to B. The main 
\OIII\,   emission features follow the equivalent H$\alpha$ structures though at typically a third to a half of their native integrated pixel intensities. 

The mid panel is the matching CTIO continuum subtracted \OIII\,   image (narrow on and off-band \OIII\,   images used). 
Both these images are presented at 90\% linear scaling to reveal the full extent of the emission 
(this sets the upper and lower limits based on the 90\% pixel intensity level where a histogram of the data is created and the limits are set to display the percentage 
about the mean value). With full pixel range, saturated pixels from the 7.7th magnitude star HD\,150915  would  prevent detail being seen.  
Low level diffuse \OIII\,   flux is seen for component\,A but it is not conspicuous. This shows it has almost been completely cancelled out indicating component\,A is mostly 
composed of normal starlight. 

The right hand panel is the off-band \OIII\,   filter image (5300\AA) that effectively represents the $V$-band starlight without the \OIII\,  +H$\beta$ emission line 
contribution. This shows dust lanes along and at the south-west extremity of the disk, concentrated starlight from components\,A, B and C and an envelope of diffuse 
starlight around the entire system. This ``common envelope" was already noted by Laustsen et al. (1987) from old $B$-band photographic data taken with the ESO 3.6-m 
telescope in the 1980s. This raises an interesting question and also a possible explanation for why the apparent axial ratios of the highly elongated component\,A differs 
so much from the surrounding oval ring. Currently known collisional ring systems with low impact parameter have axial ratios that  are broadly similar to that
of the target galaxy. Such arrangements are a natural consequence of the copious gas and ISM being swept up and compressed by the resultant density wave having a
similar radial (and perhaps also vertical) distribution as the target galaxy stellar content. This does not appear to be the case here unless what we are actually seeing 
is the residual, less-inclined bar of a more face-on disk system that extends out to the entire ring and whose presence can still be seen in the low-level diffuse  
starlight refereed to earlier. If component\,A is not a residual bar but a distinct late-type disk galaxy (as commonly assumed) stripped of gas, then the gas distribution 
could have become less flattened somehow.  Alternatively the original target galaxy could have been stretched out along the direction of the impact towards component\,B before, 
during or even after the bulls-eye collision occurred. Though component\,B is only 13$\%$ of the mass (in stars) of component\,A these effects could have been strengthened  
due to the low `Magellanic'-type masses of the system as a whole.

\begin{figure*}
\includegraphics[width=16cm,height=15cm]{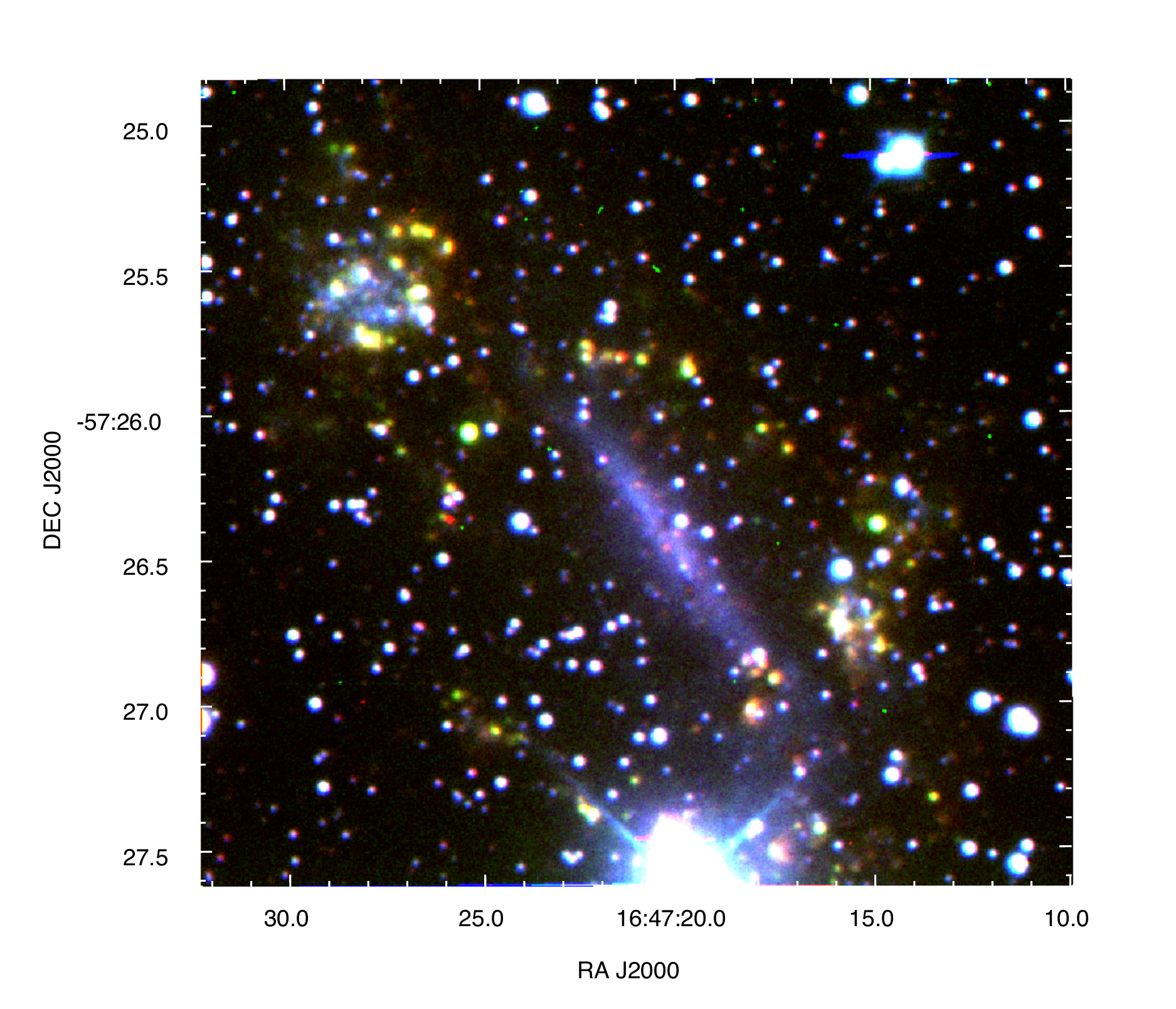}
\caption{A colour composite RGB CTIO image displaying the system's spectacular nature. The red channel is H$\alpha$, the green
\OIII\,   and the blue is broadband V. Component\,A is mostly normal starlight  with dust lanes also evident in this higher resolution
data as well as the likely effect of dust on the small H{\sc ii} knots along the disk.}
\label{CTIO-RGB}
\end{figure*}

Finally, Fig. \ref {CTIO-RGB} is a composite Red, Green Blue (RGB) CTIO image displaying the system's spectacular nature. The red channel is H$\alpha$, the green \OIII\,   and the blue 
broadband V.  Component\,A is mostly star-light with dust lanes evident in this higher resolution data. The few small H{\sc ii} knots appear red, likely due to dust 
extinction within the disk.

Based on the SHS data, literature data and our new CTIO images, Table \ref{fund_data} presents some key parameters for galaxy components A, B and C, including the 
central position of the best fit ellipse to the main emission ring. The position for component\,A is tied to the 2MASS (Skrutskie et al. 2006) near-IR imagery and 
astrometry.  The positions refer to the centre of each assumed galaxy. The centre of component\,A is derived from the dominant starlight in the $R$-band. The 
separation between the centres of components\,A and B is $\sim$80\,arcsec. Using the distance of 10\,Mpc gives a projected separation between them of 3.9\,kpc.
 
Component\,A has a major-axis of $\sim$86\,arcsec (4.2\,kpc) measured to the outer edge in the stellar continuum from off-band \OIII\,   and $V$-band 
CTIO images where a decline in the background is seen before the star-forming ring is encountered. When checked against the broad band $R$, $I$ and 2MASS images 
when autocut at the 90\% pixel intensity level the disk appears to extend only to 70-80\,arcsec.  At low isophotal thresholds starlight extends out to the star-forming ring both along the 
major and minor axes. Component\,B has a major axis of $\sim$33\,arcsec (1.6\,kpc) determined in the same way.  The strongest H$\alpha$ emission in B is also in 
an oval structure of nucleated knots $\sim$32\,arcsec across.  This is in excellent agreement with the value estimated from the concentrated starlight.  Starlight internal 
to this emission oval is clear in Fig. \ref {CTIO-RGB} and B may extend to 70\,arcsec ($\sim$3.4\,kpc) in low level, isolated clumps of star formation and diffuse 
starlight.  

The overall system envelope, including all assumed associated broad and narrow band emission and diffuse starlight as seen across the region in the right panel of Fig. 
\ref{CTIO}, has a major diameter of 318\,arcsec and a minor axis of 198\,arcsec, giving a total system size of $\sim$15.4$\times$9.6\,kpc.
 
\begin{table*}
\caption{Fundamental parameters of Kathryn's wheel. Component\,A is the main system near the ring's centre. Component\,B is the smaller system
directly to the north-east. Component\,C is projected internal to the ring to the south-west. It has also has a concentrated stellar 
component so is also considered a distinct galaxy. The oval ring completely encompasses component\,A.}
\begin{flushleft}
\begin{center}
\begin{tabular}{lccccc}
\hline
\noalign{\smallskip}
  Parameter         &     ~~~~A~~~~    &   ~~~~B~~~~  & ~~~~C~~~~ & ~~Ring~~ & ~~A-B separation~~\\
\hline
\noalign{\smallskip}
RA  (J2000)  &      16 47 19.7
                          & 16 47 27.8
                          & 16 47 15.5
                          & 16 47 21.1  
                          & ...\\
DEC  (J2000)  &   $-$57 26 31
                          & $-$57 25 35
                          & $-$57 26 44
                          & $-$57 26 36   
                          & ...\\
Size (arcsec)     &          86      &   33  &  16      & 127 & 80\\
Size (kpc)          &        4.2       &   1.6  &   0.8     & 6.2& 3.9\\
\hline
\end{tabular}
\end{center}
\end{flushleft}
\label{fund_data}
\end{table*}

\subsection{Near and mid infrared imaging}
\begin{figure*}
\resizebox{\hsize}{!}{\includegraphics{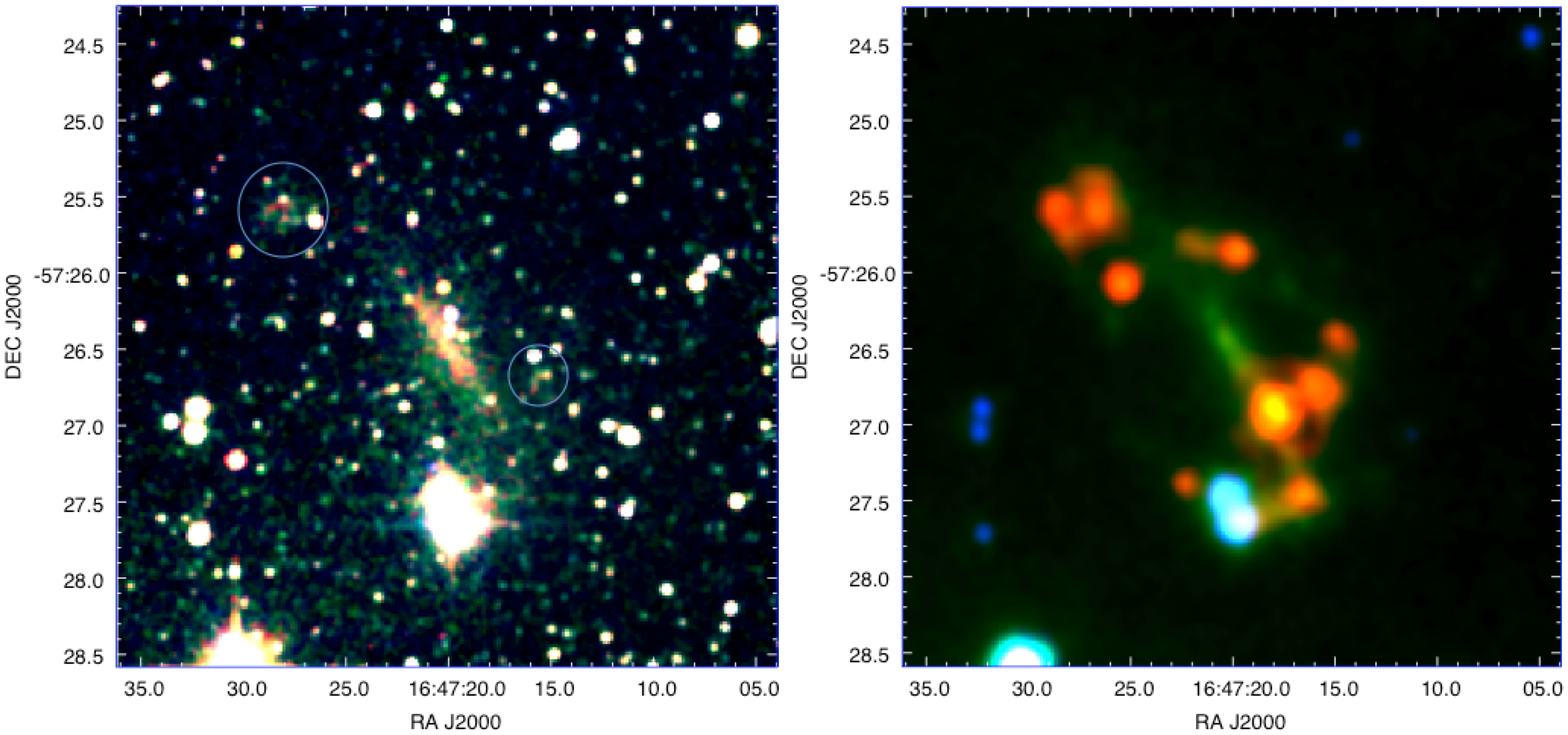} }
\caption{Left: a 4$\times$4 arcmin colour mosaic of the 2MASS $J$, $H$ and $K_s$ band data of the system (as blue, green, red). North is top and east left. 
Component\,A at the centre is clearly visible  revealing the old residual stellar population. Component\,B (surrounded by a blue circle) is obvious towards the top left of the 
image, indicating it is a distinct galaxy. Component\,C (surrounded by a blue circle to the right of component\,A) is also just visible. Right: The equivalent `hi-res' WISE mid-
IR combined colour image of bands 4, 3 and 2 (as blue, green, red) from drizzled data. Component\,A is strong in W3 centred at 12$\mu$m. The emission is likely thermal 
from diffuse interstellar dust and PAHs. Several knots in oval ring are particularly bright in W4. This is related to hot ($\sim$100\,K) dust continuum emission from 
ongoing star formation in these regions }
\label{NIR-MIR}
\end{figure*}

The 2MASS colour combined $J$, $H$ and $K_s$ band image (as red, green, blue) is presented  in the left panel of Fig. \ref{NIR-MIR}. The old stellar population of 
component\,A is clear in the near-IR and seen to be well separated  from component\,B (surrounded by a blue circle) located towards the top left of the 
image, indicating it is a distinct galaxy. Component\,C (surrounded by a blue circle to the right of Component\,A) is also just visible.
Each near-IR component is scaled at the 95\% level in the linear combination.  A low level background is also apparent which needs to be included in any overall near-IR magnitude estimates. 
The NIR magnitude differences between components A and B is $\Delta\,J\sim \Delta\,K\sim$3.1. There is also only a mild colour term for each main 
component  with J$-$K$\sim$0.3 for component A and $-$0.05$\pm$0.05 for component B.

The Wide-field Infrared Survey Explorer, WISE (Wright et al. 2010) surveyed the entire sky at 3.4, 4.6, 12 and 22$\mu$m  (bands named W1 to W4).
The right panel of Fig. \ref{NIR-MIR} shows the WISE `hi-res' mid-IR combined colour image of  bands W4, W3 
and W2 (as blue, green, red) from drizzled data.  The WISE mission provides 
``Atlas'' images, co-added from individual frames, through the web server hosted at IPAC \footnote[1]{http://irsa.ipac.caltech.edu/applications/wise/}. The images have 
beam sizes of 8.1, 8.8, 11.0, and 17.5\,arcsec for the  W1, W2, W3 and W4 bands, respectively.   

To regain valuable spatial resolution we made use of the ICORE (Image Co-addition with Optional Resolution Enhancement \footnote[2]{http://web.ipac.caltech.edu/staff/
fmasci/home/icore.html}) software (Masci 2013) which exploits the stable point spread function (PSF) of WISE. The ``drizzled" co-added images make use of a VPLR 
(Variable Pixel Linear Reconstruction) technique employing a ``Tophat" PRF (Point Response Function) kernel. The ``HiRes" images additionally use a MCM (Maximum 
Correlation Method) technique as outlined in Masci \& Fowler (2009). The drizzle images have a resolution of 5.9, 6.5, 7.0 and 12.4\,arcsec for bands W1 to W4 
respectively (Jarrett et al. 2012).  Component\,A is prominent in W3 centred at 12$\mu$m  but with sharp cut-on and cut-offs at 7 and 17 $\mu$m.  This is likely 
indicative of  thermal emission from diffuse interstellar dust and polycyclic  aromatic hydrocarbons (PAHs) in the disk.  The presence of clear dust lanes visible in the 
CTIO $V$-band image along component\,A supports this and shows that component\,A has not been entirely  stripped of its gaseous component.
 
\subsection{Existing radio data}
\begin{figure*}
\includegraphics[height=5.8cm]{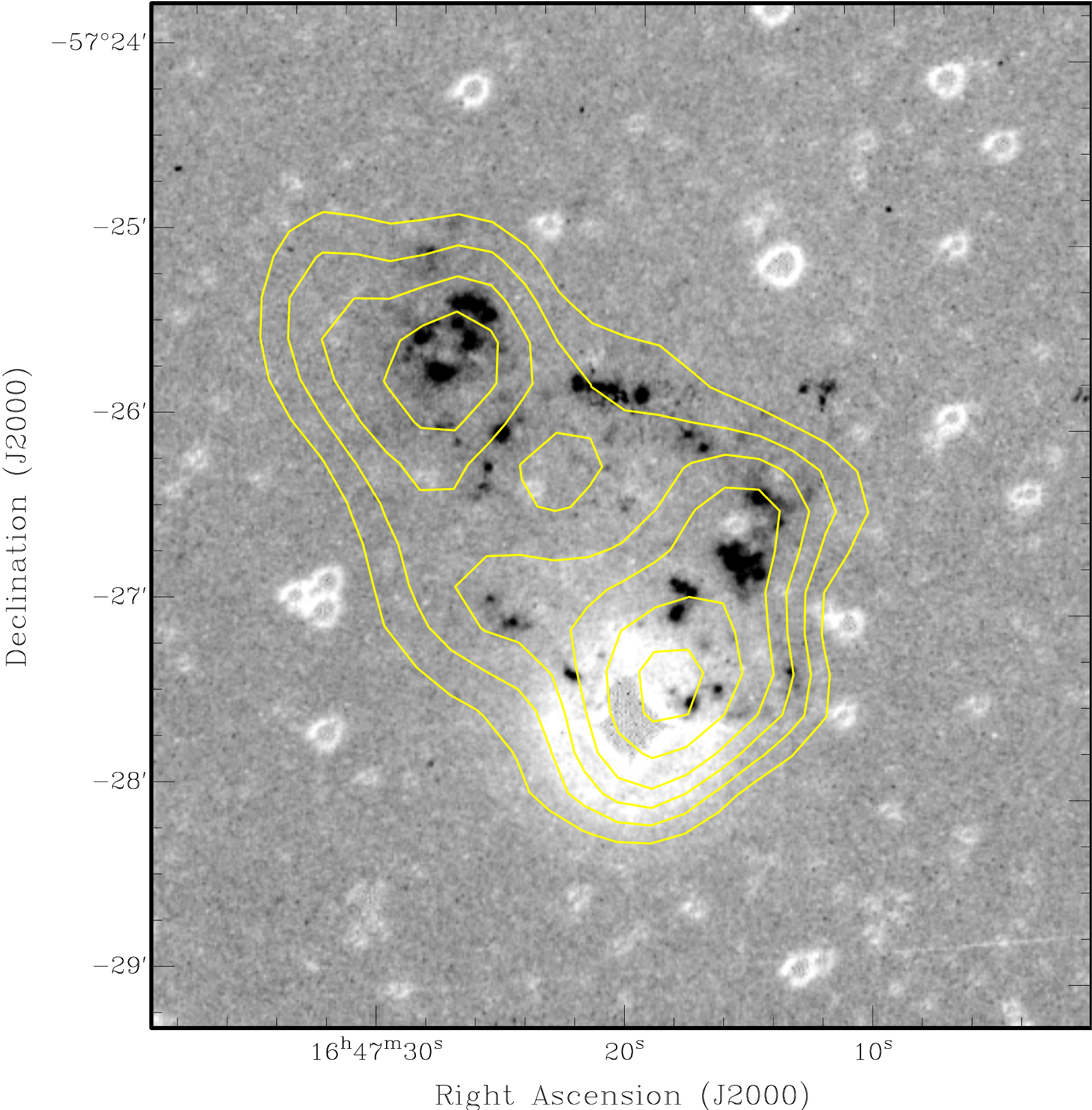}
\includegraphics[height=5.8cm]{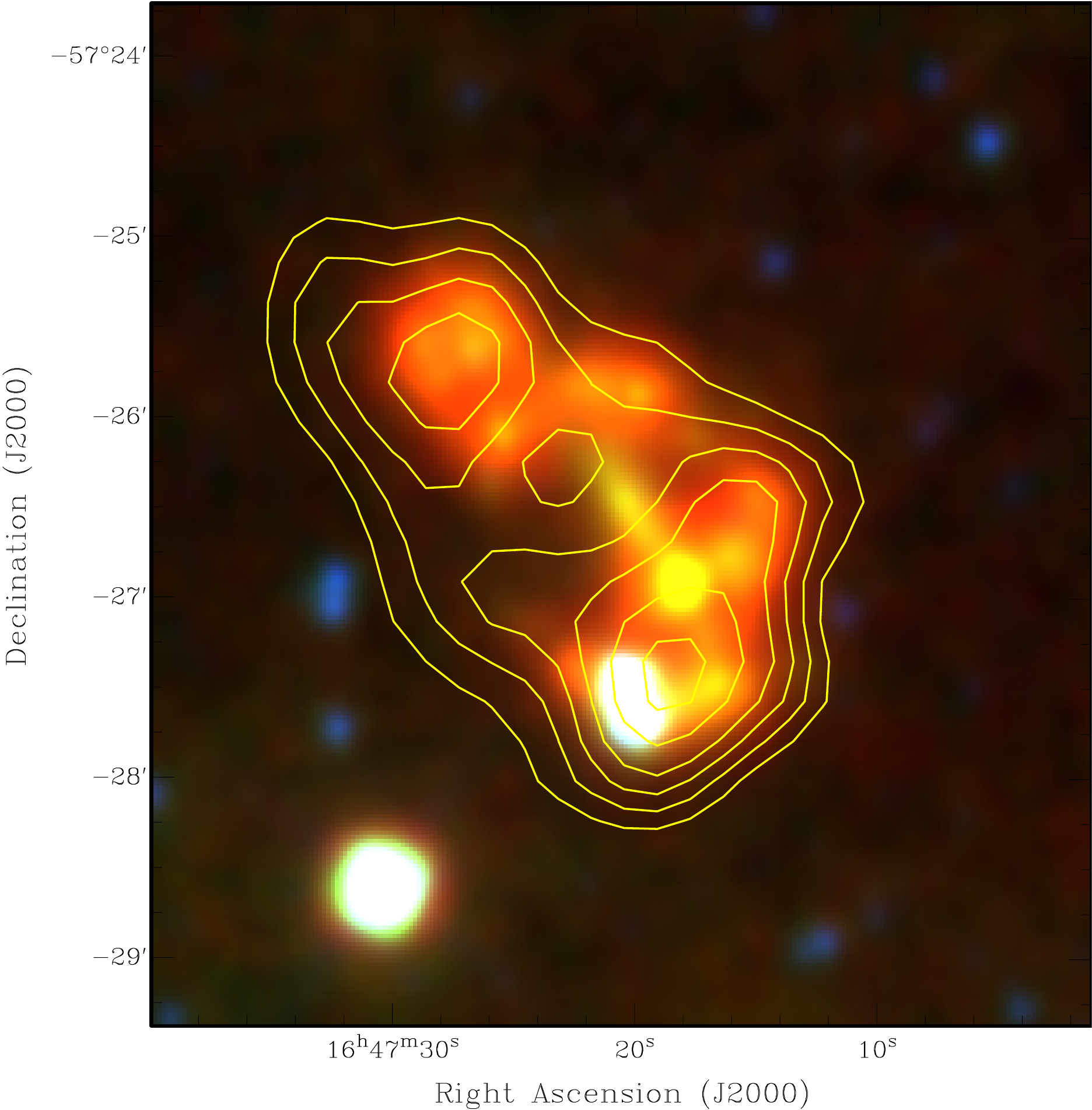}
\includegraphics[height=5.8cm]{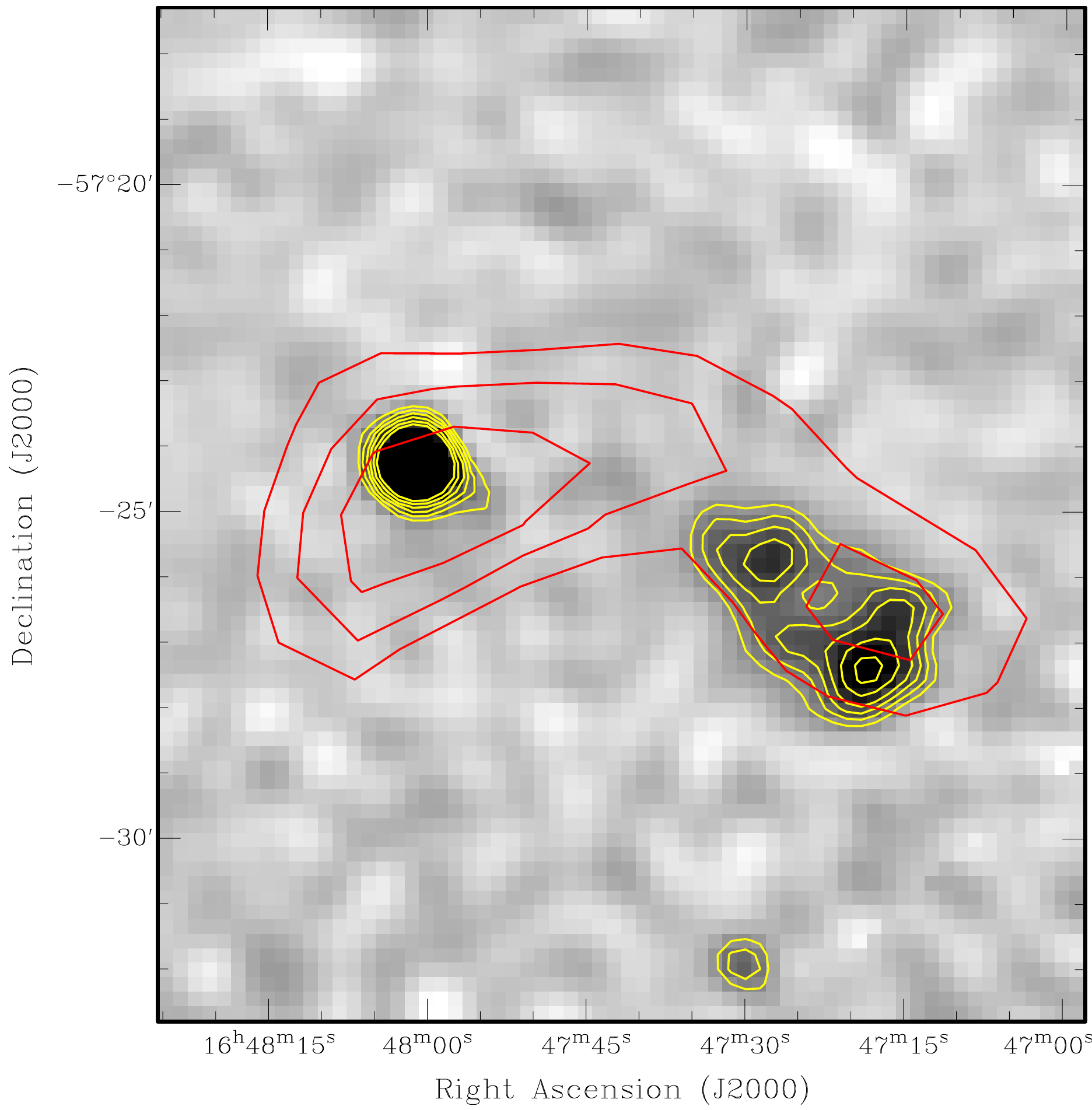}
\caption{Left and middle: SHS H$\alpha$ quotient image and WISE RGB colour composite, respectively, overlaid with contours from the MGPS2 radio 
image at 3, 5, 7, 9, 12 and 15 $\times\sigma_{mgps2}$  ($\sigma_{mgps2}=1.1$~mJy/beam).  These panels are 5$\times$6~arcmin in size. Right: MGPS2 total intensity 
image overlaid with radio contours from MGPS2 (same as 
in previous images)  and from PMN at 3, 4 and 5 $\times\sigma_{pmn}$ ($\sigma_{pmn}=8$~mJy/beam). Here the image is 13$\times$18~arcmin. North-east is to the 
to left in all panels.}
\label{radio}
\end{figure*}

In the radio regime the system is detected in the second epoch Molonglo Galactic Plane Survey (MGPS2; Murphy et al. 2007).  The MGPS2 was performed with the 
Molonglo Observatory Synthesis Telescope at a frequency of 843 MHz and with a restoring beam of $\sim$45\,arcsec.  In MGPS2, the system is resolved into two main 
components catalogued as  
MGPS~J164718-572710  (associated with component\,A) and MGPS~J164728-572544 (component\,B).  The reported peak and integrated flux densities are 15.4$\pm
$1.1~mJy/beam and 62$\pm$6~mJy for the component\,A and 11.5$\pm$1.1~mJy/beam and 34$\pm$5~mJy for component\,B. These radio counterparts are partially 
resolved, as seen in Fig. \ref{radio}. Component\,B shows a small north-east extension which doesn't correspond to either the optical or  MIR emission. Component\,A 
peaks in the region obscured by the bright foreground star HD\,150915 and also shows two fainter extensions  corresponding to the emission in both the optical and  
MIR (Fig. \ref{radio}, left and middle).  No significant radio emission at 843~MHz is observed from the 
inner region of the star forming ring associated with  component\,A. It is possible that component\,C also has a radio contribution hinted at by the overlaying elongated 
MGPS2 radio contours in this region.

The system is also detected in the Parkes-MIT-NRAO (PMN) South Survey (PMN; Wright, Griffith, Burke and Ekers 1994) at 4.85~GHz, and catalogued as 
PMN~J1647-5726.  Because of the much poorer PMN resolution ($\sim$4.2~arcmin) the system is only detected as an unresolved point source with a flux density of 
37$\pm$8~mJy, very close to the detection limit of 32~mJy. Fig. \ref{radio} (right) shows the PMN radio contours with a peak flux of 37~Jy/beam. An apparent extension 
to north-east, similar to the one at 0.843~GHz, is visible. Due to the large  beam size of the PMN survey, it is likely that this extension is caused by the confusion with the 
nearby radio point source PMN~J1648-5724 $\sim$5.5\,arcmin to the north-east.  This is  intriguing as the MGPS2 contours at the north-east edge of 
component\,B and the south-west edge around the nominal radio point source PMN~J1648-5724 are almost aligned. However, there is no evidence of an optical galaxy 
counterpart to this compact source.

Using the integrated flux densities at 4.8~GHz and 0.843~GHz (summed components\,A and B) we estimate a system-wide spectral index of $\alpha\sim$-0.5 (defined 
from $S_{\nu}\propto\nu^{\alpha}$). This estimate is uncertain for several reasons: (i) the accuracy of the flux density at 4.8~GHz is poor because of possible confusion 
with the nearby source and proximity to the sensitivity limit, (ii) the accuracy of the flux density at 0.843~GHz suffers from the measurement method (Gaussian fitting) 
which assumes that both radio objects are unresolved and (iii) the estimate of the spectral index from only two points assumes that the spectral energy distribution 
(SED) is flat in that spectral region (e.g. the free-free emission in the $\nu<1$~GHz region could be optically thick). However, the estimated radio SED is in agreement 
with SED's of normal galaxies (Condon 1992); i.e. a mixture of thermal emission originating from bright, star forming, H{\sc ii} regions, and non-thermal emission from 
diffused, ultra-relativistic electrons and supernova remnants (SNRs). 

A proper interpretation of the components's radio SED and separation of the thermal and non-thermal component (which could lead to an independent estimate of the 
SFR) requires new, high resolution and high sensitivity observations at several frequencies and in the high frequency range (1--10~GHz).

\section{Follow-up optical spectroscopy}
\begin{figure}
\resizebox{\hsize}{!}{\includegraphics{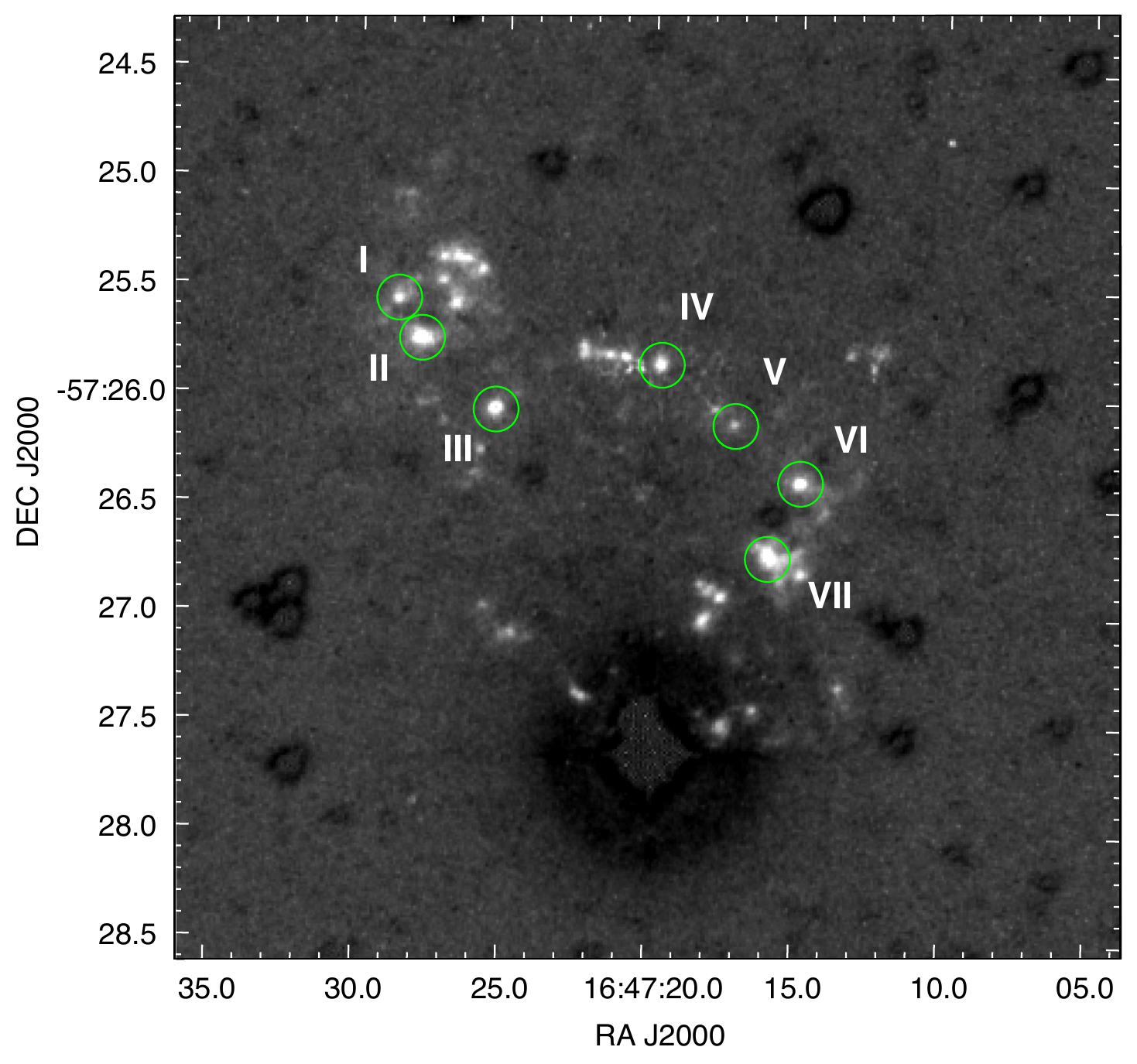} }
\caption{The locations of the spectroscopically observed emission
  knots given in Table.\ref{DBS-spec} overlaid on the
  4$\times$4\,arcminute SHS H$\alpha$ quotient image.}
\label{KNOTS}
\end{figure}

Initial spectroscopy was obtained using the now decommissioned Double Beam Spectrograph (DBS) on the 2.3\,m MSSSO telescope in June 2005. The four
positions were centred on the brightest knots in the system and are summarised in Table \ref{DBS-spec} and shown in Fig.\ref{KNOTS}. The DBS gives two 
simultaneous blue and red spectra and our choice of gratings covered the range 3520--5430\AA\ in the blue and 5950--6900\AA\ in the red. The dispersion 
was 1\AA\ per pixel in the blue arm and 0.5\AA\ per pixel in the red arm.  The 1-D spectra were obtained by summing over the extent of the compact emission on the slit 
and were flux-calibrated following standard IRAF routines. The typical on-site seeing is 1.5--2.5\,arcsec. 
 
Additional spectroscopy was obtained using the WiFes integral-field instrument on the 2.3m MSSSO telescope (Dopita et al. 2010) in July 2011. WiFes is an `image 
slicing' integral-field spectrograph made of up 25 1\,arcsecwide adjacent `slices' on the sky each of which is 38\,arcsec long. This creates an effective on-sky field of 
view of of 25$\times$38\,arcsec with the long axis oriented east-west in standard configuration.  The sky background was subtracted using empty areas of the observed 
field or from offset sky pointings. The weather was partly cloudy and though flux calibration was performed, the atmospheric transmission changed significantly between 
pointings. We used the $R=7000$ grating with coverage from 5800\AA\ to 7000\AA\ and dispersion of 0.44\AA\ per pixel. Only the prominent knots in each WiFeS 
pointing were extracted and used as the S/N in the inter-knot regions is too low (partially because of the non-optimal observing conditions)
for additional useful spectra to be extracted or to create a 2-D line intensity map or a reliable 2-D velocity field. Deeper Integral Field Unit (IFU) observations on a larger telescope are required.
 
A total of seven combined DBS and WiFeS pointings were obtained, centred on the brighter knots in the system as shown in Fig. \ref{KNOTS}, including three WiFeS 
repeats of some of the earlier DBS observations for consistency checks.  The positions are listed in Table.\ref{DBS-spec}. For WiFeS several knots can be observed 
simultaneously for each pointing.  A spectrum through the data-cube was obtained by placing apertures over each knot and wavelength summing the pixels within each 
aperture.  A combined 1-D spectrum was produced for each knot and used to derive velocities from fits to the \NII, H$\alpha$ and \SII\, emission lines and taking the 
weighted average.
 
\begin{figure*}
\begin{center}
\includegraphics[width=14cm]{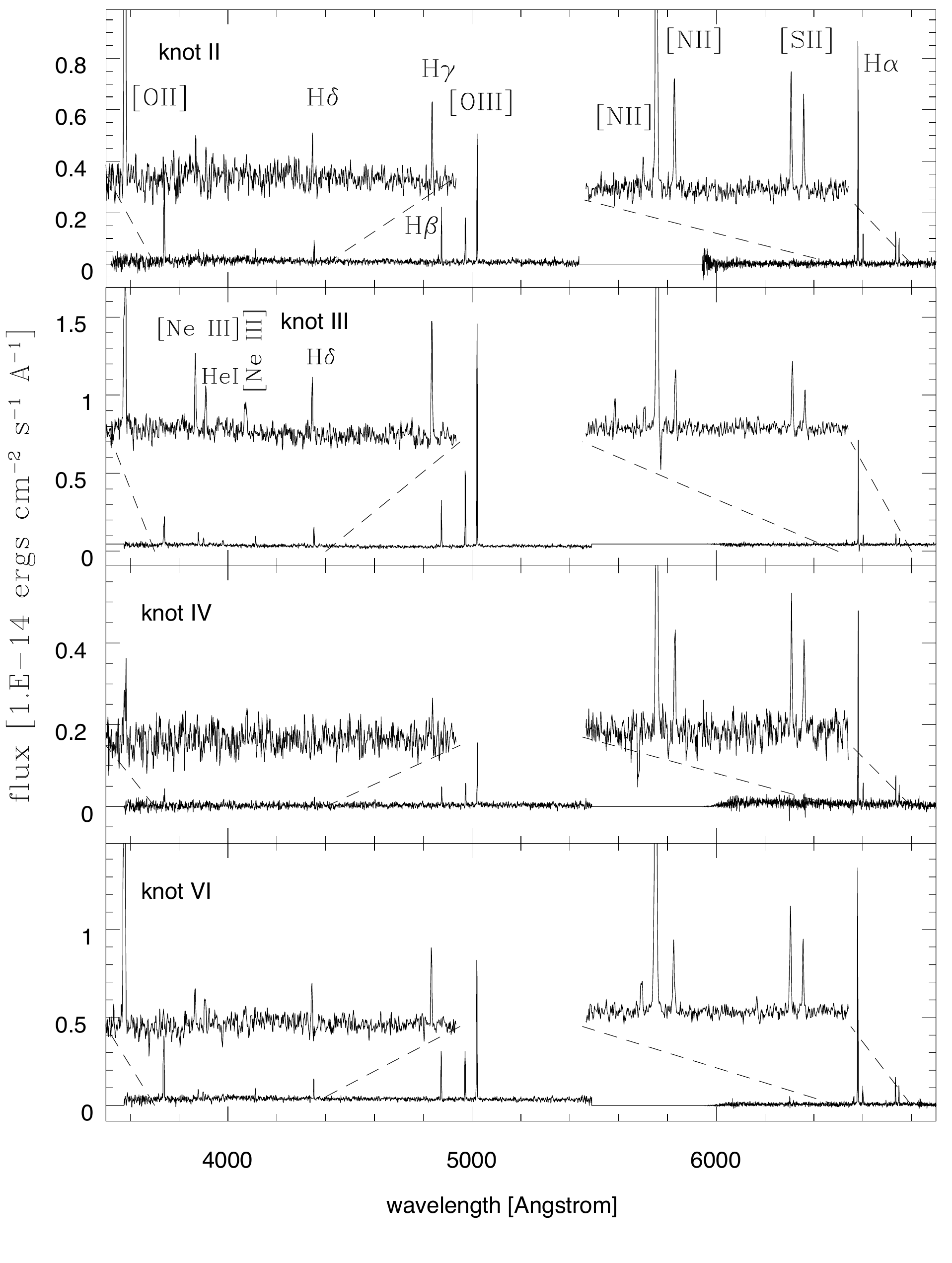}
\caption{The four DBS spectra, numbered according to position in Table \ref{DBS-spec}. Extracted spectral regions are shown enlarged to bring out fainter lines. 
The red and blue zoom windows show the spectral ranges 3700--4400\AA\ and 6500--7700\AA. The spectra are typical of H{\sc i}I regions with strong Balmer emission 
and forbidden lines of \OIII\,  , [N\,II] and [S\,II]. Note the [Ne\,III] lines in the blue data of the higher S/N spectrum for knot III.}
\label{DBS}
\end{center}
\end{figure*}

Velocities to better than $\sim$50\,km\,s$^{-1}$ are achievable with the DBS system and grating combination, but with some flop and flexure being an issue with 
this large rotating spectrograph off the Nasmyth focus (Rodgers et al. 1988).  Spectral calibration errors from the red WiFeS datacube are of the order of 0.05\AA\, (see  
Childress et al. 2014) or 2.3\,km\,s$^{-1}$ at H$\alpha$.  Arc calibrations exposures were taken for each object observation  to remove any effect of  temperature 
variations inside WiFeS.   This set-up enables us to resolve velocity dispersions across resolved systems down to $\sigma\sim$20\,km\,s$^{-1}$ (Vogt et al. 2015).   In 
our data velocity measurements of individual knots in one WiFes pointing agree to better than 10\,km\,s$^{-1}$.  DBS velocities show significant scatter of 50\,km\,s
$^{-1}$ between the blue and the red arm and are less accurate.  Where the same knot is measured with both WiFes and DBS there is acceptable agreement in two 
cases but  the offset is fairly large ($\sim$50\,km\,s$^{-1}$  in the third case (Knot~IV) and is at the upper error level that might be expected. It is clear from 
Fig. \ref{DBS} that the spectral continuum for Knot~IV is lower than for the other spectra shown.

Fig. \ref{DBS} presents the 1-D blue and red emission line spectra from the DBS for knots\,II, III, IV and V.  Extracted regions are shown enlarged to show the fainter 
lines. The spectra are typical of H{\sc ii} regions and exhibit the standard emission lines of \OII\,3727\AA, the Balmer series (H$\delta$ H$\gamma$, H$\beta$) and 
\OIII\, in the blue region. In the red there is H$\alpha$ and the weak bracketing \NII\, lines. The \SII\, doublet is also well detected and resolved providing an opportunity 
to estimate electron densities. These are near the low density limit ($n_e\sim$2--100\,cm$^{-3}$) from measurements across four knots. For knot\,III, which is the 
highest surface brightness compact emission region in the entire system, the higher S/N spectrum reveals [Ne\,{\sc iii}] lines at 3869 and 3968\AA\, rest wavelengths.  There 
is no evidence for Wolf-Rayet features in any of these H\,{\sc ii} region spectra.

\begin{table}
\caption{Summary DBS and WiFeS IFU observations from June 2005 (DBS) and July 2011 (WiFeS). 
The DBS velocities are accurate  to $\sim50$\,km\,s$^{-1}$ and WiFeS $\sim20$\,km\,s$^{-1}$}
\begin{flushleft}
\begin{tabular}{lllllrllllll}
\hline
\noalign{\smallskip}
Pointing &  RA\ (J2000) & DEC\ (J2000) &  Instrument & Velocity \\
\noalign{\smallskip}
\hline
 &   &  &   & km\,s$^{-1}$\\
\noalign{\smallskip}
Knot I & \\
  I.1  &   16:47:28.68  & $-$57:25:34 & WiFes &  799 \\
  I.2  &   16:47:26.61  & $-$57:25:34 & WiFes &  812 \\
  I.3  &   16:47:27.15  & $-$57:25:29 & WiFes &  808 \\
  I.4  &   16:47:26.60  & $-$57:25:21 & WiFes &  807 \\
Knot II & \\
 II.1  &  16:47:27.90 & -57:25:45  & DBS   & 803 \\ 
 II.1  &  16:47:27.90 & -57:25:45  & WiFes & 799 \\
 II.2  &  16:47:28.90 & -57:25:33  & WiFes & 800 \\
Knot III & \\
 III.1  &  16:47:25.30 & $-$57:26:04  & DBS   & 806 \\ 
 III.1  &  16:47:25.30 & $-$57:26:04  & WiFes & 823 \\
 III.2  &  16:47:25.80 & $-$57:26:15  & WiFes & 836 \\
 III.3  &  16:47:27.10 & $-$57:26:07  & WiFes & 834 \\
 Knot IV & \\
 IV.1  &  16:47:19.68 & $-$57:25:51  & DBS  &  827 \\ 
 IV.1  &  16:47:19.68 & $-$57:25:51  & WiFes & 880 \\
 IV.2  &  16:47:21.50 & $-$57:25:48  & WiFes & 864 \\
 IV.3  &  16:47:20.90 & $-$57:25:49  & WiFes & 871 \\
 Knot V & \\
 V.1    &  16:47:17.18 & $-$57:26:07  & WiFes & 909 \\
 V.2    &  16:47:17.82 & $-$57:26:03  & WiFes & 901 \\
 Knot VI & \\
  VI.1   &  16:47:14.80 & $-$57:26:23 &  DBS & 751 \\ 
 Knot VII & \\
 VII.1  &  16:47:15.90 & $-$57:26:44  & WiFeS & 938 \\
 VII.2  &  16:47:18.25 & $-$57:26:51  & WiFes & 910 \\
 VII.3  &  16:47:14.83 & $-$57:26:48  & WiFes & 959 \\
 VII.4  &  16:47:17.55 & $-$57:26:55  & WiFes & 932 \\
\noalign{\smallskip}
\hline
\noalign{\smallskip}
\end{tabular}
\end{flushleft}
\label{DBS-spec}
\end{table}
 
\section{Analysis}

\subsection{Optical Photometry}
Component\,A (the main ESO\,179-13 galaxy) has optical photographic photometry in the $B_{J}$, $R$ and $I$-bands (Doyle et al, 2005) while Lee et al. 
(2011) give $B \simeq$15.  In the broad-band SuperCOSMOS photographic data (Hambly et al. 2001a) the isophotal image parameters for  component\,A are blended 
with the bright star in the $B_{J}$-band so no reliable magnitude is possible from the standard image-analysis mode (IAM) processing. This 
led Doyle et al. (2005) to estimate individual Sextractor magnitudes for component\,A when constructing optical photometry for HIPASS galaxies. These were 
bootstrapped to the various SuperCOSMOS zero-points for each band. However, the areal profile fitting and image deblending techniques employed by the 
SuperCOSMOS IAM software (Hambly et al. 2001a) do enable component magnitude estimates to be obtained. There are three $R$-band magnitudes available from 
the first and second epoch broad-band surveys and from the SR exposure taken as  part of the H$\alpha$ survey.  $I$-band SuperCOSMOS magnitudes are also 
available. For isolated sources such magnitudes are accurate to $\pm$0.1.  These results are presented in Table \ref{PHOT}. The IAM results from components\,B and 
C are likely to be more reliable than for component\, A whose $I$-band magnitude in particular seems grossly  underestimated when compared to the later 2MASS near-
IR photometry and any sensible colour term. 

\begin{table}
\caption{SuperCOSMOS and Sextractor isophotal magnitude estimates for main system components. The asterisked I band entry for component\,A came from the 
IAM data from the SR/H$\alpha$ SuperCOSMOS images. $^{1}$\,Sextractor values are from Doyle et al. (2005).} 
\begin{flushleft}
\begin{tabular}{llllll}
\hline
\noalign{\smallskip}
Component &  B${_j}$ & R1 & R2 &  I  & Origin \\
\hline
A                  & 13.1 & -  & 13.3 & 12.8 & Sextractor$^{1}$\\
\hline
A                   & -  & 14.98 & 13.85 & 11.3*  & IAM (SuperCOS)\\
B                  & -  & 15.61 & 15.26 & 13.86* & IAM (SuperCOS)\\
C                  & - & 16.27 & 16.26 & 16.0 & IAM (SuperCOS)\\
\noalign{\smallskip}
\hline
\noalign{\smallskip}
\end{tabular}
\end{flushleft}
\label{PHOT}
\end{table}

\subsection{Infrared Photometry}
The available infrared photometry obtained is presented in Table \ref{ir_data} with data gathered from various archives including Skrutskie et al. (2006) for 2MASS.  The 
integrated $K_s$-band photometry was obtained from the 2MASS `Large Galaxy Atlas' (LGA: Jarrett et al. (2003).  A total $K_s$-band magnitude of 8.95 is obtained 
from an aperture of $177.8\times 56.9$\,arcsec 
that combines the three systems though the contribution from component\,C and diffuse component is minimal. We carried out differential aperture photometry on the 
two main sub-systems A and B and found a $K_s$-band difference of 2.2~mag with component A being the brighter. We find $m_K(A) = 10.3$~mag and $m_K(B) = 12.5$~mag, for 
the main bodies of components\,A and B only (i.e. the main disk for A and the core region of B).  
 
The system corresponds to IRAS 16430$-$5721 in the Point Source Catalog for the Infrared Astronomical Satellite (Beichman et al 1988).  The galaxies were detected 
in the 60 and 100\micron\ bands, but only upper limits are reported in the 12 and 25\micron\ bands.  

Both the AKARI infrared camera (IRC) All-Sky Survey Point Source Catalogue (Ishihara et al. 2010) and the AKARI far infrared surveyor (FIS) Bright Source Catalogue 
(Yamamura et al. 2009) also report flux 
densities for the objects. However, the AKARI 9\micron\ detection comes from the bright star and not the galaxy. HD\,150915 is a visual binary with 
components10\,arcsec apart. The companion star (not in SIMBAD) is redder with ($J$-$K_s$) = 1.1.  The two stars have similar magnitude in the mid-infrared: 7.45 and 
7.36 in W2, and  7.20 and 7.22 in W3. The AKARI 9\micron\ position is in between the two stars. The two stars have no infrared excess (McDonald, Zijlstra \& Boyer 2012).   The 
AKARI FIS measurements at 65 and 160\micron\ are $<3\sigma$ detections.  

The Planck Catalog of Compact Sources Release 1 (PLANCK collaboration: Ade et al. 2014) lists flux densities at 545 and 857\,GHz (550 and 350\micron\ ), although 
the 545\,GHz measurement is only a $\sim$2$\sigma$ detection. A summary of the available infrared photometry is presented in Table \ref{ir_data}.
 
\begin{table}
\caption{Infrared photometry for the combined system unless otherwise stated. WISE values are aperture corrected.}
\begin{flushleft}
\begin{tabular}{lllllrllllll}
\hline
\noalign{\smallskip}
$\lambda$  & Band & Flux density [mJy]  & [mag]  \\
\hline
\noalign{\smallskip}
1.2\micron & 2MASS J & & 9.571\\
1.7\micron & 2MASS H & & 9.122 \\
2.2\micron    &  2MASS Ks LGA   &  184     & 8.895 \\
3.4\micron    &    WISE W1     &  129.51  & 8.45 \\
4.6\micron    &    WISE  W2    &   82.38  & 8.30 \\
12\micron     &    WISE W3     &  269.80  & 5.17 \\
22\micron     &    WISE W4     &  739.20  & 2.63 \\
\\
12\micron &   IRAS 12\micron &    $<250$ \\
25\micron &   IRAS 25\micron &    $<295$ \\
60\micron &   IRAS 60\micron &    $3930 \pm 830$ \\
100\micron &  IRAS 100\micron &   $14200 \pm 1600$ \\
\\
65\micron &   AKARI FIS N60 &     $2179 \pm 748$ \\
90\micron &   AKARI FIS WIDE-S &  $4982 \pm 272$ \\
140\micron &  AKARI FIS WIDE-L &  $8730 \pm 1140$ \\
160\micron &  AKARI FIS N160 &    $6105 \pm 3460$ \\
350\micron &  Planck 857~GHz &    $7264 \pm 1629$ \\
550\micron &  Planck 545~GHz &    $1512 \pm 760$ \\
\noalign{\smallskip}
\hline
\noalign{\smallskip}
\end{tabular}
\end{flushleft}
\label{ir_data}
\end{table}
 
\subsection{H{\sc i} and optical spectroscopic velocities}

\begin{figure}
\includegraphics[width=\columnwidth,clip=true]{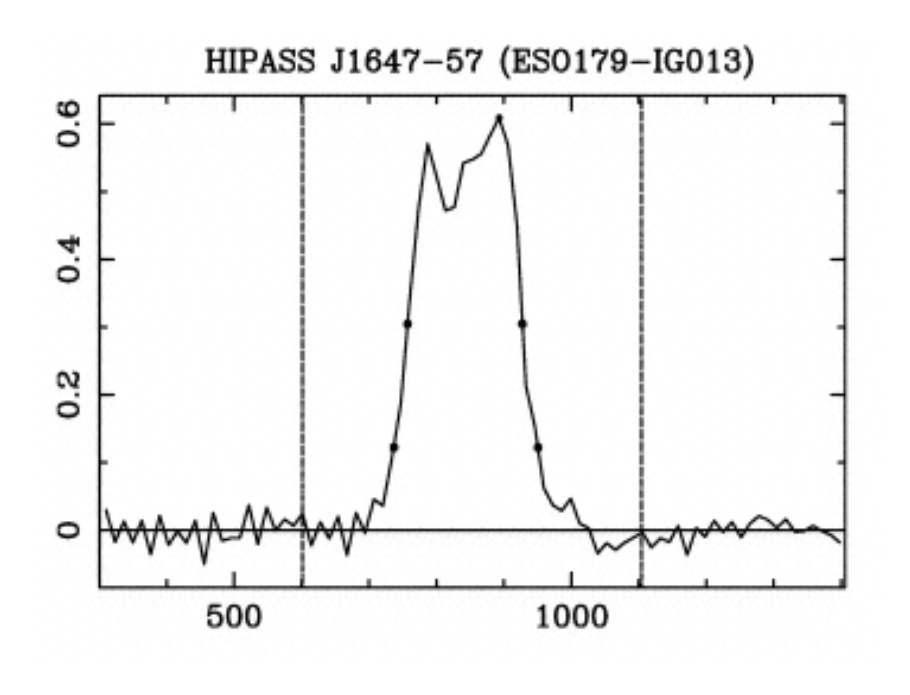}
\caption{\label{HIPASS} The H{\sc i} spectrum from the HIPASS survey (Koribalski et al. 2004) taken from a pointing at 16h47m20.0s, -57$^{o}$27'06'' (J2000). The y-axis is flux in Jy, and the x-axis heliocentric velocity in 
km\,s$^{-1}$.  The H{\sc i} velocity is 842\,km\,s$^{-1}$ taken at the midpoint of the H{\sc i} profile at the 20\% level which has a velocity width of 222\,km\,s$^{-1}$.}
\end{figure}

The H{\sc i} HIPASS spectrum of the system from a pointing at 16h47m20.0s, -57$^{o}$27'06'' (J2000) is reproduced in Fig. \ref{HIPASS}. 
The 20\%\ line width is 222\,km\,s$^{-1}$ centred at a systemic heliocentric velocity  of 842\,km\,s$^{-1}$ (Koribalski et al. 2004).    
 
For our optical observations we used the WiFeS and DBS spectra to derive heliocentric velocities of several of the most prominent emission knots, using weighted 
averages from fits to the key emission lines.  These results can be used to trace the velocity of the knots in the main ring and for the other star forming regions in 
component\,B, but not of the central disk where there is little star formation and where we did not obtain spectra. 
We rely on the published results for component\,A.  

\begin{figure}
\includegraphics[width=\columnwidth,clip=true]{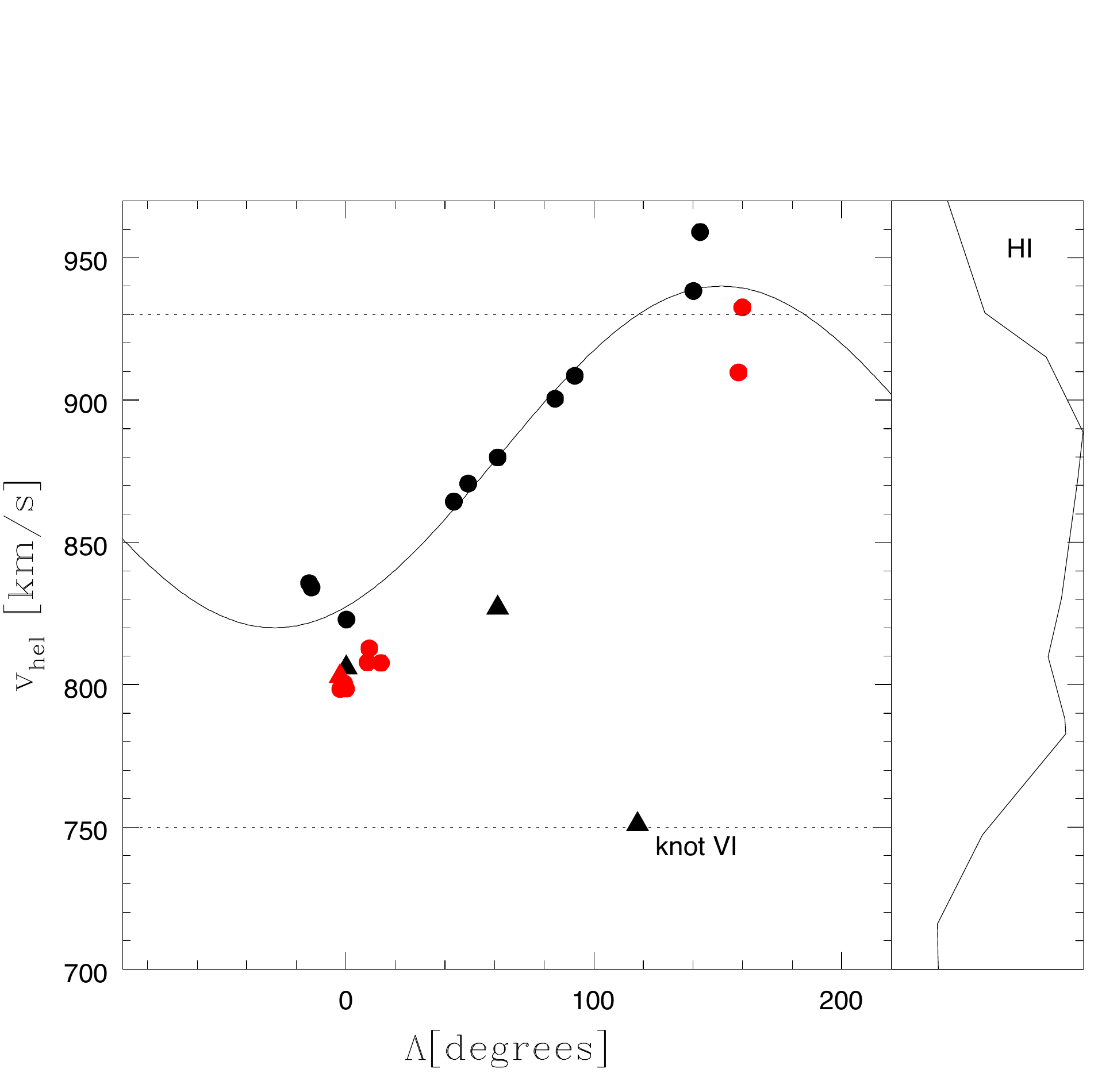}
\caption{The distribution of velocities of the star forming regions that are located along the ring (black) and further out or internal to the ring, including from component\,B and C (red).  
Circles are WiFeS data and triangles DBS. The horizontal axis shows the angle along the ring, starting from the direction of the assumed bullet, 
galaxy component\,B.On the RHS the HIPASS spectrum has been plotted for ease of comparison.}
 \label{VEL}
\end{figure}

Our new spectroscopic results are shown in Fig. \ref{VEL}. The vertical axis shows the heliocentric velocity of each knot. On the RHS of the figure we have added the 
HIPASS spectrum plotted with velocity on the same scale for ease of comparison.
The horizontal axis shows the angle along the ring, where a position angle of zero corresponds to the direction of galaxy component\,B.  This is considered the bullet because of the lack 
of other obvious galaxies in the broad-band imaging within 30\,arcmin ($\sim$5.2\,Mpc).  Black symbols are knots which we consider as part of the ring as they fall along the ring's locus even if they have a
discrepant velocity. Red symbols are for knots in galaxy components\,B or C, located inside or outside of the ring.  Circles are the WiFeS data and triangles the (less accurate) DBS data. 
There appears to be one particularly anomalous  DBS velocity for knot\,VI at only 750\,km\,s$^{-1}$. This knot is the second most intense star-forming region in the entire system. There is no 
corroborating WiFeS spectra for this knot to check the DBS result.
 
The velocity distribution for the knot samples around the ring shows a sinusoidal pattern as function of ring position angle, with systemic velocity of 880 km\,s
$^{-1}$ and an amplitude of 60\,km\,s$^{-1}$. This velocity pattern can interpreted as rotation or expansion, or a combination of both. The fact that the extrema of 
the velocity fall along the major axis is indicative of rotation being the dominant component. The velocity corrected for inclination of the ring gives 85\,km\,s$^{-1}$.  If this is 
rotation, the interior mass would be $\sim 10^{10}$\,M$_\odot$, a little higher than derived above for component\,A.

The red points at angle zero show the velocity distribution of the most conspicuous knots seen in the component\,B galaxy (the assumed bullet) 
including for knots labelled I and II in Fig.\ref{KNOTS}. The average of these is 805\,km\,s$^{-1}$, with little dispersion and is separated from that of the ring at this  
position angle of $\sim$830\,km\,s$^{-1}$.
 
\subsection{Masses}
 
\subsubsection{Derived stellar mass}

The system's stellar mass can be derived from the WISE drizzle images using the relation of Cluver et al. (2014), calibrated from the stellar masses in the 
GAMA survey (Taylor et al. 2011). The relation is:
\begin{equation}
 \log \frac{M_\ast}{L_{\rm W1}} = -2.54\left(W1 - W2  \right) - 0.17
\end{equation}

where $L_{\rm W1}$ is the in-band luminosity, i.e. the luminosity of the source as measured relative to the Sun in the W1 band as defined in
equation (1) in Cluver et al. (2014).

Using  $L_{\rm W1} = 8.27 \times 10^9\rm \, L_\odot $, we obtain a stellar mass of $M_\ast = 2.4\times 10^9$\,M$_\odot$ for the entire system. The colour correction used in this method aims at accounting for the contribution from a young stellar population. Without this correction derived masses would have been a few times higher.

The main galaxy component stellar masses can also be estimated from their $K_s$-band luminosities.  The $K_s$-band magnitudes were measured using 2MASS aperture 
photometry, subtracting the field stars within the aperture. The $K_s$-band image (Fig.\ref{NIR-MIR}) clearly shows the main galaxy 
and a weak but definite detection of galaxy component\,B.  The star forming ring is not discernible but evidence of a faint wash of near-IR flux internal to the ring is seen 
as well as a weak detection interior to the south-west edge of the ring associated with component\,C previously identified as WKK\,7457. As the $K_s$-band primarily 
shows the older stellar population, the main body of component\,A appears most prominent.

Using the 2MASS $K_s$ imaging and assuming a mass-to-light ratio of 0.80 (Bell et al. 2003), we find a stellar mass of the main components $1.29\times 10^9\,\rm M_\odot$
and $1.70\times 10^8\, \rm M_\odot$ for  A and B, respectively.  Their mass ratio is a factor of 7.5.

The integrated $K_s$-magnitude of 8.95 gives a total stellar mass of $4.7\times 10^9\,\rm M_\odot$, in broad agreement with the WISE derivation. The mass-to-light 
ratio at $K_s$ varies by a factor of two over a wide range of stellar ages excluding the very youngest stars (Drory et al. 2004). The models of Henriques 
et al. (2011) show that the $K_s$-band can be significantly enhanced by younger, more massive, AGB stars, and include a component from red supergiants.
 
\subsubsection{Gas and dust mass}
 
The H{\sc i} neutral hydrogen mass is reported by Koribalski et al. (2004) as $M_{HI} = 2 \times10^9\,\rm M_\odot$. This is similar to the overall stellar mass of the entire 
system which is therefore gas-rich. It is not clear where the bulk of the gas originates but the large H{\sc i} mass makes an original association with component\,A more 
likely.

The mass in molecular gas is not known. CO data is required to determine this directly. Stark et al. (2013), provide evaluations of  molecular to neutral 
Hydrogen (H\,2/H{\sc i}) ratios from a broad sample of field galaxies spanning early-to-late types, evolutionary stage and stellar masses of 10$^{7.2}$ 
to 10$^{12}\,\rm M_\odot$. Their sample is not statistically  representative because CO measurements for low-mass systems such as ours are lacking. It does seem that 
for a range in $g-r$ of $\sim$0.3, molecular to neutral gas ratios range from 1 to $\sim$10 percent so $\sim10^9\,\rm M_\odot$ in molecular hydrogen is possible.

We use two different methods to estimate dust masses using available flux estimates.  
The AKARI 65, 90, 140 and 165\micron\ flux densities were measured using methods optimized for unresolved sources (Yamamura et al. 2009). Since the system is 
larger than the AKARI beam, the AKARI measurements may miss some of the extended emission\footnote{any contributions from the bright stars to the south 
are negligible, e.g. the SED for HD\,150915 star drops by $>$3 orders of magnitude between 1 and 10$\mu$m.} and could refer mainly to component\,A. This could 
explain why the AKARI flux estimates are lower by a factor of two compared to the equivalent IRAS values.  
We present AKARI fluxes for completeness but only use the WISE, IRAS, and {\it Planck} data in this 
analysis. 

The first method used fits the infrared SED with a blackbody function multiplied by an emissivity function that scales as $\lambda^{-\beta}$.  We then 
used the resulting temperature $T$ in the expression:
\begin{equation}
M_{\rm dust}=\frac{f_\nu D^2}{\kappa_\nu B_\nu (T)},
\end{equation}
\noindent where $f_\nu$ is a flux density measured in a single wave band, $D$ is the distance to the source, $\kappa_\nu$ is the wavelength-dependent opacity, and $B_\nu (T)$ is the blackbody function.
 
We perform fits to the data using $\beta$ values of 1.5, as found empirically by Boselli et al. (2012) using {\it Herschel} data, and 2, which corresponds to the theoretical 
emissivity function given by Li \& Draine (2001) that is frequently used in dust mass calculations.  We also use the $\kappa_\nu$ based on values from Li \& Draine 
(2001) to calculate the dust mass corresponding to the modified blackbody fits with $\beta=2$.  For other $\beta$ values it is common to rescale $\kappa_\nu$ by $
(\lambda/\lambda_0)^{-\beta}$, where $\lambda_0$ is some reference wavelength where the emissivity is  known.  Though many authors select 250\,$\mu$m as a 
reference wavelength, it is unclear if this is appropriate, as the choice is usually poorly justified.  We only calculate dust masses for fits where $\beta=2$, but still report 
the temperatures for both $\beta=1.5$ and $\beta=2$.   The dust masses from our fits are derived from the amplitudes of the best fitting modified blackbody functions.  
We use a function for $\kappa_\nu$ where $\lambda_0$=240~$\mu$m and $\kappa_\nu$(250~$\mu$m)=5.13 cm$^2$ g$^{-1}$.
 
Fits were performed to the 100, 350, and 550\micron\ data that tends to originate from the coldest and most massive dust component heated by the diffuse interstellar 
radiation field. Emission at shorter wavelengths may come from a warmer thermal component (Bendo et al. 2006, 2012, 2015).  The 100\micron\ emission may also originate 
in part or entirely from dust
heated by star formation.  We need the data point to constrain the Wien side of the spectral energy SED, but we are also aware that the temperature may be 
overestimated by our analysis and the dust mass underestimated.
 
We calculate the dust mass using the flux density at 350~$\mu$m,  the longest wavelength at which we have a $>3 \sigma$ detection. At shorter wavelengths,  small 
uncertainties in the temperature will result in relatively large uncertainties in the dust mass.  We adopt $\kappa_\nu(350\mu\mbox{m}) = 2.43$\,g\,cm$^{-2}$ for the 
opacity.
 
We obtain dust temperatures of $19 \pm 2$\,K for fits with $\beta=2$ and $23 \pm 3$~K for $\beta=1.5$.  The slope of the 350--550\micron\ data is more 
consistent with a $\beta=2$ emissivity function than the $\beta=1.5$ emissivity function found by Boselli et al. (2012), but given the large uncertainty in the 550\micron\ 
data, it would be difficult to rule out the possibility that $\beta$ is smaller than or larger than 2.  The dust mass for the fit corresponding to $\beta=2$ is $(1.1 \pm 
0.3)\times10^7$ M$_\odot$.
 
In the second method, we fit the model templates of Draine et al. (2007) to the data.  These templates are calculated for several different environments (the Milky Way, 
the LMC, and the SMC), different PAH mass fractions and different illuminating radiation fields.  Draine et al. (2007) suggest fitting spectral energy distribution data using 
the following equation which we also adopt:
\begin{equation}
f_\nu = A \left[ (1-\gamma)j_\nu(U_{\rm MIN})
 + \gamma j_\nu(U_{\rm MIN}, U_{\rm MAX}) \right]
\label{DraineFitEq}
\end{equation}

In this equation, $j_\nu(U_{\rm MIN})$ is the template for dust heated by a single radiation field $U_{\rm MIN}$ and is equivalent to cirrus heated by a diffuse interstellar 
radiation field.  The quantity $j_\nu(U_{\rm MIN}, U_{\rm MAX})$ is the template for dust heated by a range of illuminating radiation fields ranging from $U_{\rm MIN}$ to 
$U_{\rm MAX}$, with the relation between the amount of dust $dM$ heated by the a given radiation field in the range between $U$ and $U+dU$ given by the equation:
\begin{equation}
{\rm d}M \propto U^{-2}\, {\rm d}U
\end{equation}

This $j_\nu(U_{\rm MIN}, U_{\rm MAX})$ effectively represents the fraction of dust heated by photodissociation regions (PDRs) in the fit and the $\gamma$ term 
indicates the relative fraction of dust heated by the PDRs.  The $A$ term is a constant that scales the sum of the $j_\nu$ functions (which are in units of Jy\,cm$^2$\,sr
$^{-1}$\,H-nucleon$^{-1}$) to the flux density.  It is related to the dust mass by:
\begin{equation}
M_{\rm dust}=\frac{M_{\rm dust}}{M_{\rm gas}} m_{\rm p} D^2 A
\end{equation}
where $M_{\rm dust}/M_{\rm gas}$ is the dust-to-gas ratio, $m_{\rm p}$ is the mass of the proton, and $D$ is the distance.
 
As this system shares many characteristics of dwarf galaxies and Magellanic systems (including the metallicity -- see below), we only fit LMC and SMC templates to the 
data.  We use all WISE, IRAS, and {\em Planck} data between 12 and 550\micron\ in the fit.
 
We obtained best fits using templates based on the LMC2\_05 model from Draine et al. (2007). The resulting $U_{\rm MIN}$  and $U_{\rm MAX}$ were 1.5 and 
$10^6$ times the local interstellar radiation field. The best fitting $\gamma$ was $0.024 \pm 0.017$.  This is consistent with the majority or even all the dust mass being 
contained in a cirrus component.  The resulting dust mass is $(1.5\pm0.3)\times10^7$ M$_\odot$, slightly higher than that calculated using the single 
modified blackbody fit. This is expected given that the Draine et al. (2007) models  account for the mass in warmer dust components and for the relative contributions of 
various thermal components of dust to  the SED. The masses are summarised in Table \ref{masses}. Individual stellar masses of the two systems are $\sim$3 times less than that of 
the total system, so much of the  stellar mass is located outside of the brighter regions. Deeper mid-IR images to find the detailed location of the remainder, either in a halo or in 
expelled streams, would  be of interest.
 
 The two galaxy components\,A and B are small systems. For comparison, the stellar mass of the LMC is $M_\ast(LMC) = 2.7 \times 10^9\,\rm M_\odot$. Its W1 flux is 
2080\,Jy (Melbourne \& Boyer, 2013) which scaled to 10~Mpc gives 50\,mJy, about three times fainter than Kathryn's Wheel.  Component\,A is  similar to the 
LMC.  Component\,B is comparable to the SMC, which has a stellar mass of $ 3.1\times10^8\,\rm M_\odot$.
  
The system contains more gas and dust than the LMC and SMC.  The LMC H{\sc i} mass  is only $0.48 \times 10^9\,\rm M_\odot$ (Staveley-Smith et al. 2003), 
and the SMC has an H{\sc i} mass of $5.6 \times 10^8\,\rm M_\odot$ (Van der Marel, Kallivayalil \& Besla 2009). While the stellar masses are broadly comparable, the 
dust mass is three times higher and the neutral gas four times higher than that of the LMC and SMC combined.  We do not know with certainty which of the two 
components (A or B) is the origin of the gas and dust emission. The gas-to-dust ratio is $\sim$133, which excludes the unknown molecular mass and is therefore an 
upper limit.  For comparison, a mean gas-to-dust ratio of 58.2 was found for 78 bright galaxies in the Herschel Virgo Cluster Survey (Davies et al. 2012). Based on their 
study they suggest that a factor of $\sim2.9$ can be used to convert atomic gas to total gas (i.e. H{\sc i} + H\,2 ) in their survey.  For comparison, the value of the gas-to-
dust ratio for the Milky Way is $\sim$143 (Draine et al. 2007), a similar value to what we derive here.  Interestingly, the ratio of the extent of the star forming regions within B 
to that of the ring around A is similar to the ratio of the galaxy stellar masses.
\subsection{Star formation rates}
 
Star formation rates (SFR) can be derived from different tracers (e.g. Kennicutt \& Evans 2012), mostly related to the UV photons produced by young massive stars.  The H$\alpha$ luminosity and the mid-IR flux tracing heated dust are independent tracers of the SFR.

\subsubsection{SFR from the H$\alpha$ Flux}

The SFR from the integrated H$\alpha$ luminosity of the system is calculated following Kennicutt \& Evans (2012), using the expression:
\begin{equation}
\log \frac{\dot M_\ast}{\rm M_\odot\,yr^{-1}} = \log
L({\rm H}\alpha) - 41.27
\label{eq:SFR}
\end{equation}

Our first estimate of the integrated H$\alpha$ flux (and luminosity) is derived from aperture photometry on all of the clearly identifiable knots in the  CTIO continuum-subtracted
H$\alpha$ image. These included 43 knots in the ring, 8 knots in the core of component\,B, and 8 knots in the region of the disk that  defines component\,A.   The
resulting fluxes were converted to individual star formation rates for each knot. 

Aperture photometry was also done on the full areas covered by each of the main system components\,A, B, C and the ring. This provided a global integrated SFR
as H$\alpha$ emission which is  distributed more widely around the entire region and even at low levels along the main disk of component\,A.  The results are
listed in Table \ref{masses}. For component\,B we measured the main body, but excluded the faint outer regions because of confusion with the ring to the
south. Table \ref{masses} also lists the specific SFR (SSFR: the SFR divided by the stellar mass).  For the disk system component\,A  and for B we used the stellar
masses given in Table \ref{masses}. No mass was derived for the ring.  For the full system we included all the stellar mass.

The conversion factor of flux units\,count$^{-1}$ was derived from the individual H{\sc ii} knots listed in Table \ref{DBS-spec}, using a 7\,arcsec aperture for each. The H$\alpha$ flux for these
knots was measured from the calibrated WiFeS IFU data. Poor weather during the WiFeS observations resulted in a factor of two spread in measured conversion
factors.  Removing the lower values left knot systems I, II and IV. Averaging their conversion factors yields $1.2 \pm 0.3 \times 10^{-18}$\,erg\,cm$^{-2}$\,s$^{-1}$
per count for the CTIO image. The photometry counts were converted to flux units, and 
a total flux for all components of $F$(H$\alpha$) = $3.0 \pm 0.6 \times 10^{-12}$\,erg\,cm$^{-2}$\,s$^{-1}$ was obtained.

Owing to possible problems with the flux calibration, an independent measurement would be useful. Unfortunately, the system is just outside the footprint of 
the VST/OmegaCAM Photometric 
H-Alpha Survey: VPHAS+, (Drew et al. 2014), and while it is imaged by the calibrated low-resolution the Southern H-Alpha Sky Survey Atlas (SHASSA) 
survey (Gaustad et al. 2001), the proximity of HD\,150915 
precludes the measurement of an accurate integrated H$\alpha$ flux following the recipe of Frew et al. (2013).  However, the SHS
(Parker et al. 2005) offers an alternative approach to obtaining the integrated H$\alpha$ flux of the system, in order to compare with our CTIO estimate.  The SHS
was absolutely calibrated by Frew et al. (2014a), who showed flux measurements to 20--50$\%$ accuracy were feasible for non-saturated diffuse emission regions. 
Using the methods discussed there, and correcting for reduced transmission due to the systemic velocity, we determine an integrated H$\alpha$ flux 
of $1.1 \pm 0.5 \times 10^{-12}$\,erg\,cm$^{-2}$\,s$^{-1}$ for component\,B through an aperture of 90\,arcsec; a correction for [N\,II] contamination is 
unnecessary based on the high excitation and low metallicity of the system's H{\sc ii}
regions.  This agrees with an equivalent CTIO measurement to within the uncertainties, so we average the results to get a total flux for component\,B.  Unfortunately the
proximity of the bright star precludes an accurate flux measurement from the SHS for the star-forming ring.  An approximate lower limit, based on a
simple curve-of-growth analysis of several apertures is $F$(H$\alpha$) = $\sim$1$\times 10^{-12}$\,erg\,cm$^{-2}$\,s$^{-1}$, also in agreement with the CTIO
measurement.  The total H$\alpha$ flux for the system, now using the weighted average of the independent CTIO and SHS calibrations, 
is $2.9 \pm 0.3 \times 10^{-12}$\,erg\,cm$^{-2}$\,s$^{-1}$, adopted hereafter.

The systemic flux, now corrected for a reddening $E(B$$-$$V)$ = 0.24 mag, and applying a Howarth (1983) reddening law, is $F_0$(H$\alpha$) = $5.1 \times 10^{-12}$\,erg\,cm$^{-2}$\,s$^{-1}$.  This leads to a total 
luminosity, $L({\rm H}\alpha)$ = $6.1 \times 10^{40}$ ergs\,s$^{-1}$, and a total ionizing flux, $Q$ = 4.4 $\times 10^{52}$ photons\,s$^{-1}$, for a distance of
10\,Mpc.   The integrated SFR was then found to be $0.33 \pm 0.09$\,M$_\odot$\,yr$^{-1}$ following equation\,\ref{eq:SFR}.  This SFR is about 60$\%$ higher than that of the LMC,
adopting the H$\alpha$ luminosity for the LMC from Kennicutt et al. (1995).  Our estimate may be an underestimate as we have not corrected for any foreground reddening, though the extinction derived from the mean 
Balmer decrement of the brightest H{\sc ii} regions from our spectra is consistent with the Schafly \& Finkbeiner (2011) extinction, indicating that the internal reddening is, in at least the most luminous H{\sc ii} regions, 
minimal.  There may also be photon leakage from the larger H{\sc ii} regions in the system as noted by Rela\~no et al. (2012), which might help to explain the difference between the SFRs derived from H$\alpha$ and 22\,
\micron\ WISE flux (see below).

We note that the most luminous H{\sc ii} region complex in ESO\,179-13 is centred on Knot VII, with $L({\rm H}\alpha) \simeq 5 \times 10^{39}$ ergs\,s$^{-1}$.  This 
luminosity is comparable to supergiant HII regions like 30 Doradus in the LMC and NGC 604 in M\,33 (Kennicutt 1984). Knot VII is within component C which contains old 
stars as it is detected in the 2MASS imagery as a separate object. It also lies interior to the fitted ring and has a different velocity distribution to the components along the 
ring, exhibiting the largest velocities across the system. We still consider it likely to be a separate star-forming dwarf galaxy but further observations are required to establish this
clearly and to study the relationship between component\,A and C.

\subsubsection{SFR from the Mid-IR Flux}

The warm dust continuum is best traced by the 22\,\micron\ WISE flux.  Cluver et al. (2014) have calibrated the relation of this flux against H$\alpha$-derived SFRs
(corrected for extinction) from the GAMA Survey (Gunawardhana et al. 2013) and find:
\begin{equation}
 \log \frac{\dot M_\ast}{\rm M_\odot\,yr^{-1}} = 0.82\,\log\,\nu L(22\,\mu{\rm m}  [L_\odot])  - 7.3.
\end{equation}
This yields an integrated SFR of 0.47\,M$_\odot$\,yr$^{-1}$ for ESO\,179-13, somewhat higher than that derived from the de-reddened H$\alpha$ flux.  In contrast, the 12\micron\
emission (Cluver et al. 2014), calibrated in the same way, corresponds to a lower SFR of 0.15\,M$_\odot$\,yr$^{-1}$.  The lower mass measured in the 12\,\micron\ band may be because this band detects significant
PAH emission.  This is relatively weak in low metallicity environments (e.g. Engelbracht et al. 2005, 2008), so emission from bands that contain PAH emission and the 
SFRs based on these bands may be underestimated.  All in all, the various estimates are in reasonable agreement, and we conclude that the integrated SFR is approximately 0.2--0.5\,M$_\odot$\,yr$^{-1}$.
 
\begin{table*}
\caption{\label{masses}Stellar, gas and dust masses, star formation rates  and specific star formation rates  estimated for the full system and, in so far as is known, for
the individual components. Values for component\,A refer to the inner region (the disk), and for the core of component\,B.  The SSFRs for A and B assume total stellar masses
of $1.3 \times 10^9$ and $1.7 \times 10^8$ M$_\odot$ respectively (refer to the text).}
\begin{center}
\begin{tabular}{llcccc}
\hline
 Parameter                            	&                                                                       	&  Full system               	&  Component A           	&  Component B           	&~~~~~Ring~~~~~	\\
\hline
$M_\ast$ ($K_s$)                     	& [M$_\odot$]                                                  	& $4.7\times 10^9$       	&  $1.3\times 10^9$      	&  $1.7\times 10^8$   	&       ...      		\\   
$M_\ast$ (WISE)                     	& [M$_\odot$]                                                    	& $2.4\times 10^9$        	& ...                                   	&   ...                        		&       ...            	 	\\
$M_{\rm HI}$                            	& [M$_\odot$]                                                    	& $2 \times 10^9$          	& ...                                   	&  ...                            	&       ...            	 	\\
$M_{\rm dust}$                        	& [M$_\odot$]                                                    	& $1.5\times10^7$         	&  ...                                  	&  ...                         		&       ...             	\\
SFR (H$\alpha$)                    	& [M$_\odot$\,yr$^{-1}$]                                   	&  0.33                              	&  0.01                    		&  0.08                          	&     0.24               	\\    
SFR (WISE 12\,$\mu$m)         	& [M$_\odot$\,yr$^{-1}$]                                	&  0.15                           	&  ...                           	&   ...                                 	&       ...             	\\
SFR (WISE 22\,$\mu$m)         	& [M$_\odot$\,yr$^{-1}$]                                 	&  0.47                            	&  ...                             	&   ...                                 	&       ...             	\\
SSFR (H$\alpha$)                	& [M$_\odot$\,yr$^{-1}$\,M$_\ast^{-1}$]~~~~~ 	& $7.0 \times10^{-11}$      &  $7.7 \times 10^{-12}$    	&   $4.7 \times 10^{-10}$  	&        ...             	\\

\hline
\end{tabular}
\end{center}
\end{table*}

\subsubsection{Star formation distribution}
  
We find that almost all of the SFR in component\,A (the main disk or bar) is accounted for by the main identified knots. In contrast, only 28\%\ of the SFR in 
component\,B appears concentrated in the main knots. About 44\%\ of the SFR in the main ring is incorporated in the most intense and well defined knots.  The remainder 
may be located in knots below the confusion limit of the available images.

The SFR in the galactic disk that comprises component\,A is almost entirely due to the bright region towards the south west, located on the edge of the main disk and 
interior to the fitted oval of the main ring shown in Fig. \ref{CTIO}. There are a few further small, faint, reddened knots in the central regions but the remainder of the 
galaxy seems dormant. The star formation in component\,B is mainly concentrated in an oval region that is $30\times20$\,arcsec in size  that overlaps with the
underlying galaxy detected in the optical broad band colours and 2MASS as galaxy WKK\,7463. However, there is star formation throughout a region 30--40\,arcsec in 
radius centred on B with one small isolated blob even located 60\,arcsec from the core to the north-west. There is clearly a continuum of star formation over much of the 
entire area where that associated with component\,B blends in to the north-east edge of the ring.
 
\begin{figure}
\includegraphics[width=\columnwidth,clip=true]{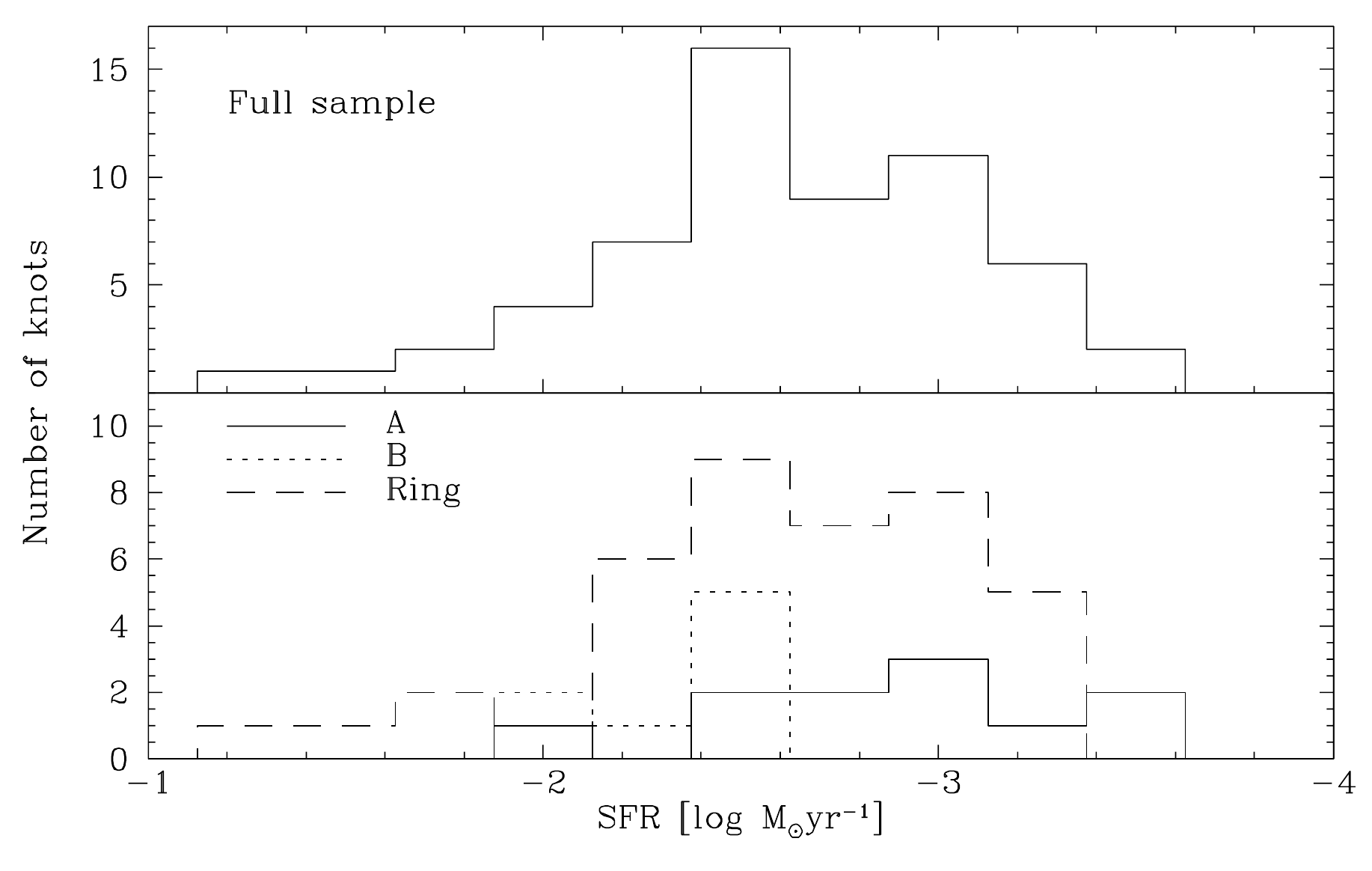}
\caption{\label{SFR} The distribution of star formation rates within the knots, for component\,A (along the disk),
component\,B (the assumed 'bullet') and the ring. The top panel shows the combined distribution. }
\end{figure}

The distribution of SFR over the detected knots is shown in Fig. \ref{SFR}. The top panel shows the sum of the knots in components\,A, B and the ring, and the 
bottom panel shows the individual components. The brightest knot in the ring (III.1) has a SFR of 0.06\,M$_\odot$\,yr$^{-1}$.  Galaxy component\,B lacks the tail 
to low SFR seen in the knots of component\,A and in the ring: this is likely due to the confusion limit being much brighter for B.  The two strongest regions of SFR are in the ring.
 
\subsubsection{Star forming versus non star forming galaxies}
 
\begin{figure} \resizebox{\hsize}{!}{\includegraphics{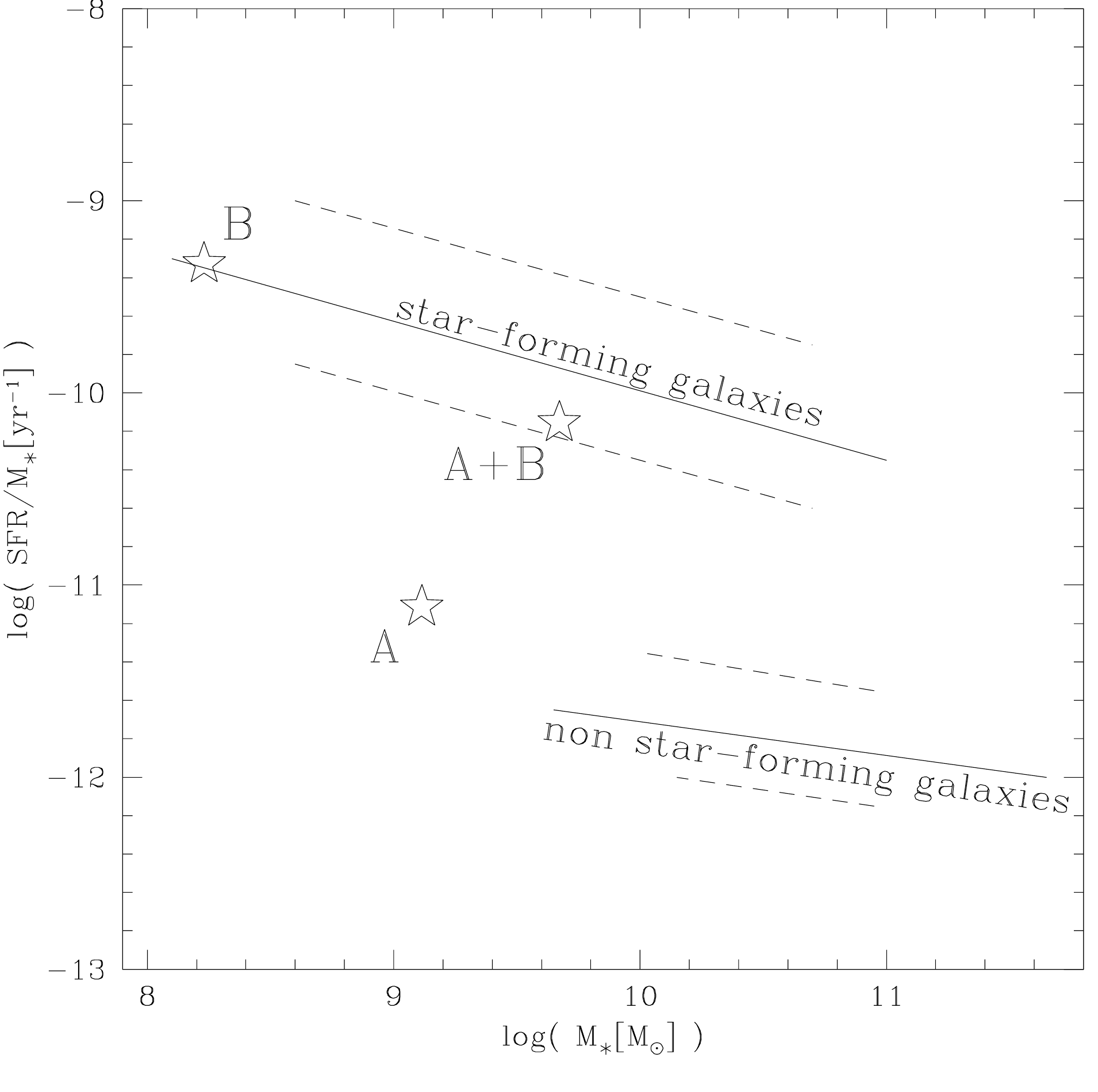} }
\caption{The distribution of star-forming and non star-forming galaxies in the stellar mass versus specific star formation plot from Schiminovich et al. (2007) and adapted 
by Kennicutt \& Evans (2012). Squares show locations of galaxies\,A, B and the full system.  The ring is not included in the SFR for component\,A but is in the 
full system.}  
\label{ssfr} \end{figure}
 
Schiminovich et al. (2007) have shown that galaxies distribute over two clear sequences when the total stellar mass is plotted against specific star formation rate. The 
approximate location of the sequences is shown in Fig. \ref{ssfr}; the dashed lines show the 1-$\sigma$ spread of the distribution. We plot the locations of component\,A 
component\,B and the total system. The SSFRs are taken from Table \ref{masses}. The masses are taken as the full stellar mass divided over components A and B 
according to the ratio of their individual stellar masses. Note that component\, A includes the full mass but its SSFR only covers the apparent disk.

Component\,B falls well above the star-forming sequence while component\,A fits close to the non-star forming galaxies as expected from a  naive view of their 
star-forming activity evident in the imagery. Component\,A would be seen as quiescent if it were not for the bright region of star-formation at the south-west 
edge of the disk.  The SSFR may be underestimated if extinction is significant.  Combining the entire system and using the full stellar mass and including the ring gives a 
data point at the lower end of the star-forming sequence. The combined system is not vigorous, even though individual components in the system are.  

\subsection{Abundances}

\begin{figure}
\resizebox{\hsize}{!}{\includegraphics{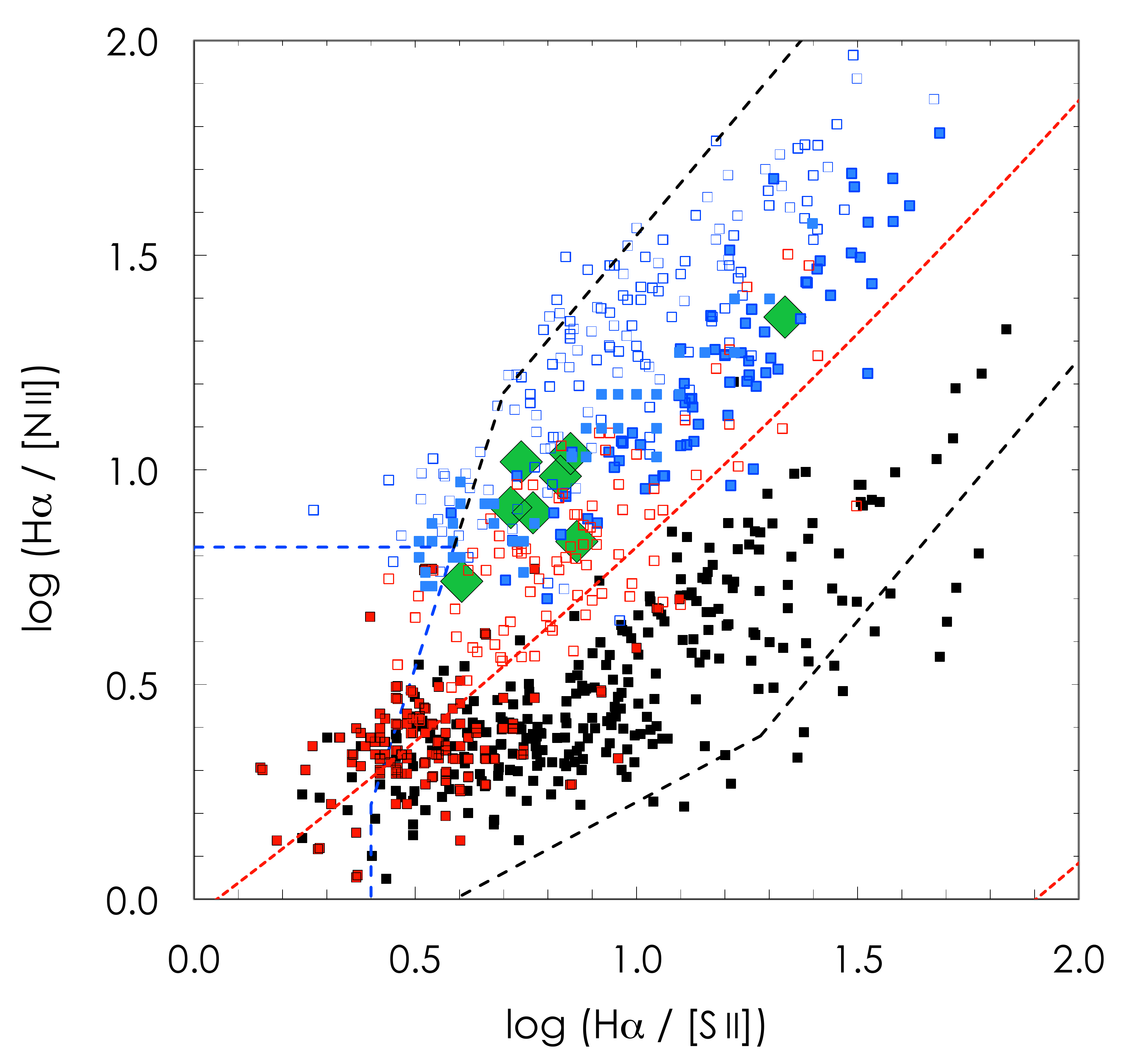} }
      \caption{Line ratios plotted for H{\sc i}I regions in several galaxies: Milky
    Way (filled black squares), M\,31 (filled red squares), M\,33 (open red
    squares), LMC (filled blue squares), other LG dwarf galaxies (open blue
    squares).  Refer to the text for further details. The ESO\,179-13 emission knots are shown by the large green diamonds. }
    \label{metallicity}
\end{figure}

The observed optical emission-line spectra of the knots can be used to derive an approximate metallicity, using the oxygen abundance as a proxy.   
The weakness of the [N\,II] lines flanking H$\alpha$ in each spectrum indicates low metallicity. Modelling of the spectra is made difficult by the lack of the \OIII\,    
4363\AA\ line which is the main electron temperature indicator. However, metallicities can also be approximated from the [N\,II] and [S\,II] lines which are well  detected. 
Fig. \ref{metallicity} plots the line intensities relative to H$\alpha$ of [N\,II] and [S\,II], where [N\,II] refers to the 6584\AA\ line (the brightest of the  doublet) while [S\,II] 
refers to the sum of both the 6713\AA\ and 6731\AA\ lines.  This plot, like the commonly used `BPT' diagram (Baldwin, Phillips \& Terlevich 1981), is used 
as a diagnostic to separate different types of  emission-line nebulae, but Viironen et al. (2007) and Frew \& Parker (2010) have recently shown its sensitivity 
as a metallicity and ionization parameter diagnostic.  This diagram also circumvents the use of the 
\OIII\,   line which is not always detected, and uses a small wavelength range in the red which improves the  relative line-to-line calibration and extinction estimates. The 
[N\,II] and [S\,II] lines are also produced at similar temperatures, alleviating the sensitivity of the collisional forbidden lines to metallicity. The plotted diagnostic is a 
combination of the `N2' and `S2' indicators of Pilyugin, Vilchez \& Thuan (2010).

Fig.\,\ref{metallicity} shows the location of spectral line ratios from H{\sc ii} regions measured across a variety of galaxies, in order of decreasing metallicity: M\,31, the 
Milky Way, M\,33, the LMC and lower metallicity dwarf systems in the Local Group.  The data are taken from Viironen et al. (2007, and references therein) and Frew et 
al. (in preparation).  The boundary lines are adopted from Frew \& Parker (2010) and Frew et al. (2014b), with the black-dashed trapezoid showing the domain of H{\sc ii} regions. The plot 
shows a clear trend with metallicity with the different systems being well separated. Note that red curves enclose the planetary nebula domain (generally of higher 
excitation) and the blue-dash enclosed area at the lower left is occupied by shock-excited supernova remnants (Sabin et al. 2013).  The filled green diamonds show the 
location of the individual knots measured in ESO\,179-13.  Their location in the plot coincides well with H{\sc ii} regions in the LMC, and therefore the galaxies can be 
considered to have a similar metallicity.  Indeed, the relations of Pilyugin et al. (2010) indicate a metallicity of [O/H]\,$\simeq-0.4$ for ESO\,179-13, consistent with a 
value for the LMC.  This makes it likely that the majority of the gas feeding the knots originated from galaxy\,A, which has a similar total mass and luminosity to the LMC.  
Component\,B has a mass similar to the SMC and may therefore be expected to have a lower metallicity and gas content.

\subsection{Ages}

When did the bull's-eye encounter occur between galaxies\,A and B?  An age estimate can be derived from the projected angular separation between the two 
components of 3.9\,kpc and the observed heliocentric velocity difference between them. We assume they are effectively at the same distance and that the observed 
difference in the heliocentric velocities is due to their peculiar motions and that there is no relative acceleration. No transverse velocity component can be obtained so 
the estimated age is a lower limit. We use the average of the three literature value for component\,A from Strauss et al. (1992), Di Nella et al., (1997) and  Woudt et al. 
(1999) of 761\,km\,s$^{-1}$ together with our newly determined value for component\,B from the average of several prominent emission knots in the system which gives 
a velocity of 806 $\pm$ 32\,km\,s$^{-1}$ from fits to 13 emission lines across the blue and red spectral regions. There is a nominal 45\,km\,s$^{-1}$ difference which 
produces an encounter age of 87\,million years. If we assume the transverse velocity is of the same order as that along the line of sight then this introduces a root 2 
factor leading to a revised age of 123\,million years. System age can also be estimated from the expansion assumed for the ring. Using an assumed inclination of 
45\,degree, and a velocity amplitude (de-projected rotational/expansion velocity) of 85\,km\,s$^{-1}$ then an age of 30\,Myr is obtained. This is a lower limit as the 
velocities are mostly due to rotation, and the above assumes all is expansion.

\section{Discussion}

\subsection{Dwarf rings}

ESO\,179-13 is among the smallest collisional ring galaxies known. The ring diameter of 6.2\,kpc is the smallest of the Madore et al. (2009) catalogue, where diameters 
range up to 70\,kpc.   Components\,A and B are separated by only 4\,kpc, while in the full catalogue, the colliding components are almost always more than 10\,kpc 
apart (Conn et al. 2011).  The absolute K-band magnitude of ESO\,179-13 is $M_K=-21.1$.  Madore et al. (2009) list only 10 objects that are less luminous. 

Because ESO\,179-13 shows a full ring it is a unique case of a dwarf cartwheel-type galaxy. It is more than 1.5\,mag (a factor of four) fainter than the next 
faintest example, Arp 147 ($M_K=-22.8$mag).  It shows that the process causing full rings can also act in interacting dwarf galaxies.  Such galaxies have smaller cross 
sections and shorter dynamical time scales, which may make such occurrences less frequent, but does not stop ring formation altogether.

The difference in $K$-band magnitude between components\,A and B is 2.2\,mag. This is at the very top end of the distribution in Madore et al. (2009).  
As component\,A is already one of the smallest galaxies in the Madore et al. sample, component\,B is possibly the smallest of all known bullet galaxies causing (partial) 
rings.  The mass ratio between A and B may be close to the limit for the formation for collisional rings formation (D'Onghia et al. 2008).

\subsection {Environment}

Collisional ring galaxies tend to be found in low-density environments and avoid dense clusters (Madore et al. 2009).  Kathryn's Wheel 
as ESO\,197-13 is also seen to be in a region devoid of major galaxies.  The SHS images show no H$\alpha$-emitting companions within a degree of ESO\,179-13. The 
broadband SuperCOSMOS  $R$-band data reveal no other obvious galaxy of any kind within half a degree of the centre of Kathryn's wheel.  There is a HIPASS galaxy 
with an almost identical velocity 1.5 degrees away: J1636-57 
but  is much fainter, with an integrated flux of 2.8 Jy.km\,s$^{-1}$.  
Given the similarity in velocity and the angular separation of 260\,kpc, a physical association is possible. 
Galaxy J1620-60 also has the same velocity, is a few times fainter (37.4 Jy.km\,s$^{-1}$) than ESO\,179-13, but is over seven degrees, over 1\,Mpc,  away.
The nearest large galaxy is NGC\,6744, a grand-design barred spiral (Ryder, Walsh \& Malin 1999). It has an identical velocity and assumed distance (Kankare et al. 
2014) and a projected separation of 19.1\,deg, or 3.2\,Mpc from ESO\,179-13.   It is unlikely to have interacted with this system.

\subsection{Impact parameter}

The ring's centre is displaced from that of of component\,A by about 580\,pc. We take this as the impact parameter of the collision with the bullet component\,B. 
The chances of a random (first) encounter between two isolated systems so close to the centre of A seems remote.  Assuming three galaxies within a sphere of 250\,kpc 
radius, and a relative velocity of 100\,km\,s$^{-1}$, the chance of such a close encounter within a Hubble time is approximately $10^{-4}$. Assuming $2\times 10^5$  
galaxies within 10\,Mpc, the chance of one such a close encounter happening within 10\,Mpc within the past $10^8$\,yr (the life time of the ring) is 15\%.    This 
calculation suggests that ESO\,197-13 is likely the closest collisional ring galaxy to the Milky Way.

The calculation ignores gravitational focussing. The gravity of the main galaxy bends the trajectory of the bullet towards it, increasing its capture cross section. This 
decreases the resultant impact parameter $b$.  For $v=100\,\rm km\,s^{-1}$ and $M = 6.6 \times 10^9\,\rm M_\odot$, the reduction is given by:
\begin{equation} 
\frac{b}{b_0} = \sqrt{\left(1 + \frac{2 G M}{b v_0^2}\right)^{-1}} \approx 0.7
\end{equation} 

This doubles the chances to 30\%, and suggests a space density of $7 \times 10^{-5}\,\rm Mpc^{-3}$.  This estimate shows that the existence of one such system within 
10\,Mpc is quite plausible. 

\subsection{Lessons and prospects}

The discovery of Kathryn's Wheel shows that the processes forming full rings can act in Magellanic type systems.  Full rings have only been seen and studied in large, 
spiral galaxies, and so their formation has been assumed to require systems with halo masses above $5\times 10^{11}\,\rm M_\odot$ (e.g. D'Onghia et al. 2008) though 
the catalogue of Madore et al. (2009) does seem to extend to a few galaxies as small as ESO\,197-13.  
As dwarf galaxies outnumber other Hubble types by a factor of $\sim1.4$ (Driver 1999), dwarf collisional rings may also outnumber those systems involving spiral galaxies.  
This is shown by the space density derived above: it is 10 times higher than that of Few \&\ Madore (1986), whose numbers are derived from more luminous galaxies.

Both components appear to have been gas-rich prior to collision. We assume the ring is formed in gas associated with component\,A, and so was gas-rich even 
though  it shows little star formation now (as is normal for residual galaxies within collisional  ring systems). 
The collision may have diverted its gas out of the main disk. It is noteworthy 
that component\,B shows vigorous star formation and may show its own (small) ring. It survived the encounter much better than A whose future evolution will  be 
dictated by the depletion of its original gas reservoir even if the underlying stellar population seems largely unperturbed.   Arp\,147 is another case where both colliding 
galaxies formed rings though they are very different in nature. Table \ref{summary_data} gives a summary table of the remaining key system parameters previously available or newly compiled.

\begin{table}
\caption{Summary table of key parameters previously available and newly derived for this system}
\begin{flushleft}
\begin{center}
\begin{tabular}{ll}
\hline
\noalign{\smallskip}
Velocity H{\sc i} (helio)   km\,s$^{-1}$        		&  $842\pm5$ 		\\   
Velocity (gal-centric)  km\,s$^{-1}$ 			&  $736\pm6$       	\\
Velocity (Local Group) km\,s$^{-1}$~~~~~~    	&  $678\pm11$    	\\
Distance (LG)   Mpc    					&  $10.0\pm0.7$   	\\
\hline
Max. system extent  (arcsec) 				&  $318\times$198   \\
Max. system extent  (kpc) 					&  $15.4\times$9.6   \\
System metallicity 	[O/H]					&  $\sim-0.4$  		\\ 
System age 		(Myr)					&  30--120 		\\
Space density 		(Mpc$^{-3}$) 			&  $7 \times 10^{-5}$	\\
\hline
\end{tabular}
\end{center}
\end{flushleft}
\label{summary_data}
\end{table}

ESO\,197-13 is an important system that can be used to study collisional rings as it is much closer than any equivalent, opening new possibilities for analysis. System 
dynamics can be studied at unprecedented spatial resolution as 0.1\,arcsec (e.g. HST and/or AO resolution) is only 25\,pc. This is sufficient to explore details of the 
system-wide  star formation process and to allow detection of individual stars.   It would be a prime candidate to measure the star formation history in the regions 
recently traversed by the ring.

\section{Summary}
During visual scans of the SHS (Parker et al. 2005), a spectacular new collisional ring galaxy (or Cartwheel analogue) has been found, ESO\,179-13, which we have 
named Kathryn's Wheel.  With an estimated distance of only 10.0\,Mpc, it is in the local supercluster but has been previously missed due its location near the Galactic 
plane and its projected proximity to a bright 7.7-magnitude star which interferes with observation. 

The velocity range of all optical emission components from the main ring, component\,B and the published velocity for component\,A is 
about 200\,km\,s$^{-1}$. This is completely encapsulated within the FWHM of 222\,km\,s$^{-1}$ reported for the H{\sc i} profile at the 20\% level centred at a 
heliocentric velocity of 842\,km\,s$^{-1}$ which we consider representative of the overall interacting system.    

At a distance of 10\,Mpc, the system has an overall physical size of $\sim$15\,kpc for all major components and the diffuse surrounding envelope, a total mass 
of $M_\ast = 6.6\times 10^9$\,M$_\odot$ (stars + H{\sc i}) and a metallicity of  [O/H]$\sim$\,-0.4, that classifies it as a Magellanic-type system. It has a large reservoir of 
neutral gas and is the 60$^{\rm th}$ brightest galaxy in the HIPASS H{\sc i} survey.  By some margin it is also the nearest ring system known. This system offers an 
unprecedented opportunity to undertake detailed studies of this rare phenomenon, where two reasonably well matched galaxies have a close encounter centred close to 
the potential well of the primary. 

\section{Acknowledgments}
This paper used the SIMBAD and VizieR services operated at the CDS, Strasbourg, the NASA/IPAC Extragalactic Data base (NED) and data 
products of the AAO/UKST H$\alpha$ Survey, produced with the support of the Anglo-Australian Telescope Board and  UK Particle Physics and Astronomy 
Research Council (now the STFC).  This research used data from the Wide-field Infrared Survey Explorer, a joint project of the University of California, Los 
Angeles, and Jet Propulsion Laboratory/California Institute of Technology, funded by the National Aeronautics and Space Administration, and data from the 2MASS, 
a joint project of the University of Massachusetts and the Infrared Processing and Analysis Center California Institute of Technology, funded by the National Aeronautics 
and Space Administration and the National Science Foundation. The authors acknowledge help of Dr Anna Kovacevic at the CTIO telescope. We are grateful for 
stimulating discussion with Kate Zijlstra. We thank an anonymous referee who provided excellent feedback to help improve the quality of this paper.
 


\begin{thebibliography}{}  
\bibitem[]{}Amram P., Mended de Oliveira C., Boulesteix J., Balkowski C., 1998, A\&A, 330, 881
\bibitem[]{}Appleton P.N., Struck-Marcell C., 1996, Fund. Cosmic Phys., 16, 111
\bibitem[]{}Arp H.C., 1966, ApJ, 140, 1617
\bibitem[]{}Arp H.C., Madore B., 1987,  A Catalogue of Southern Peculiar Galaxies and Associations. Cambridge, New York: Cambridge University Press
\bibitem[]{}Baldwin J.A., Phillips M.M., Terlevich R., 1981, PASP, 93, 5
\bibitem[]{}Barnes J.E., Hernquist L., 1992, ARA\&A, 30, 705
\bibitem[]{}Beichman C.A., Neugebauer G., Habing H.J., Clegg P.E., Chester T.J., 1988, The Infrared Astronomical Satellite (IRAS) Catalogs and Atlases. Volume 1: Explanatory supplement
\bibitem[]{}Bell E.F., McIntosh D.H., Katz N., Weinberg M.D., 2003, ApJS, 149, 289 
\bibitem[]{}Bendo G.J. et al., 2006, ApJ, 652, 283
\bibitem[]{}Bendo G.J. et al., 2012, MNRAS, 419, 1833
\bibitem[]{}Bendo G.J. et al., 2015, MNRAS, 448, 135
\bibitem[]{}Bianchi L., 2014, Ap\&SS, 354, 103B
\bibitem[]{}Block D.L. et al., 2006, Nature, 443, 832
\bibitem[]{}Boselli A. et al., 2012, A\&A, 540, 54
\bibitem[]{}Boselli A., Fossati M., Gavazzi G., Ciesla L., Buat V., Boissier S., Hughes T., 2015, A\&A, 579, 102
\bibitem[]{}Brosch N., Vaisanen P., Kniazev A.Y., Moisav A., 2015, MNRAS, in press, arXiv:1506.00409
\bibitem[]{}Burbidge E.M., Burbidge G.R., Prendergast K.H., 1964, ApJ, 140, 1617
\bibitem[]{}Buta R., 1995, ApJS, 96, 39
\bibitem[]{}Childress M. J., Vogt F. P. A., Nielsen J., Sharp R. G., 2014, Ap\&SS, 349, 617
\bibitem[]{}Cluver M. et al., 2014, ApJ, 782, 90
\bibitem[]{}Condon J.J., 1992, ARA\&A, 30, 575
\bibitem[]{}Conn B.C.,  Pasquali A., Pompei E., Lane  R.R.,  Chene A-N., Smith R., Lewis G.F., 2011, ApJ, 741, 80
\bibitem[]{}Corwin H.G., de Vaucouleurs A., de Vaucouleurs G., 1985, Univ. Texas Monogr. Astron., 4, 1
\bibitem[]{}Davies J.I. et al., 2012, MNRAS, 419 3505
\bibitem[]{}Dennefeld M., Lausten S., Materne J., 1979, A\&A, 74, 123  
\bibitem[]{}de Vaucouleurs G., de Vaucouleurs A., Corwin H.G., Buta R.J., Paturel G., Fouque P., 1991, Third Reference Catalogue of Bright Galaxies, Volume III. New York: Springer
\bibitem[]{}di Nella H., Couch W.J., Parker Q.A., Paturel G., 1997, MNRAS, 287, 472
\bibitem[]{}D'Onghia E., Mapelli M., Moore B., 2008, MNRAS, 389, 1275
\bibitem[]{}Dopita M., et al., 2010, Ap\&SS, 327, 245
\bibitem[]{}Doyle M.T., et al, 2005, MNRAS, 361, 34
\bibitem[]{}Draine B.T. et al., 2007, ApJ., 663, 866
\bibitem[]{}Drew J.E. et al., 2014, MNRAS, 440, 2036
\bibitem[]{}Driver S.P., 1999, ApJ, 526, L69
\bibitem[]{}Drory N., Bender R., Feulner G., Hopp, U., Maraston C., Snigula J., Hill, G.J., 2004, ApJ, 608, 742
\bibitem[]{}Engelbracht C., et al., 2005, ApJ, 628, L29 
\bibitem[]{}Engelbracht C., et al., 2008, ApJ, 678, 804
\bibitem[]{}Few J.M.A., Madore B.F., 1986, MNRAS, 222, 673
\bibitem[]{}Fosbury R.A.E., Hawarden T.G, 1977, MNRAS, 178, 473 
\bibitem[]{}Freeman K.C., de Vaucouleurs G., 1974, ApJ, 194, 569
\bibitem[]{}Frew D.J., Parker Q.A., 2010, PASA, 27, 129
\bibitem[]{}Frew D.J., Boji\v{c}i\'c I.S., Parker Q.A.,  2013, MNRAS, 431, 2
\bibitem[]{}Frew D.J., Boji\v{c}i\'c I.S., Parker Q.A., Pierce M.J., Gunawardhana M.L.P., Reid W.A., 2014a, MNRAS, 440, 1080 
\bibitem[]{}Frew D.J. et al., 2014b, MNRAS, 440, 1345
\bibitem[]{}Gaustad J.E., McCullough P.R., Rosing W., Van Buren D., 2001, PASP, 113, 1326
\bibitem[]{}Gerber R.A., Lamb S.A., Balsara D.S., 1992, ApJL, 399, L51    
\bibitem[]{}Gil de Paz A., Madore B.F., 2003, ApJS, 147, 29
\bibitem[]{}Graham J., 1974, Observatory, 94, 290
\bibitem[]{}Gunawardhana M.L.P. et al., 2013, MNRAS, 433, 2764
\bibitem[]{}Hambly N.C. et al., 2001a, MNRAS, 326, 1279
\bibitem[]{}Henriques B., Maraston C., Monaco P., Fontanot F., Menci N., De Lucia G., Tonini C., 2011, MNRAS, 415, 357
\bibitem[]{}Higdon J.L., Higdon S.J.U., Rand R.J., 2011, ApJ, 739, 97
\bibitem[]{}Hinshaw G. et al., 2013, ApJS, 208, 19 
\bibitem[]{}Howarth I., 1983, MNRAS, 203, 301
\bibitem[]{}Ishihara D. et al., 2010, A\&A, 514, 1
\bibitem[]{}James P.A. et al., 2004, A\&A, 414, 23
\bibitem[]{}Jarrett T.H., Chester T., Cutri R., Schneider S.E., Huchra J.P., 2003, AJ, 125, 525
\bibitem[]{}Jarrett T.H. et al., 2012, AJ, 144, 68
\bibitem[]{}Kankare E. et al., 2014, A\&A, 572, A75
\bibitem[]{}Kennicutt R.C., 1984, ApJ, 287, 116
\bibitem[]{}Kennicutt R.C., Bresolin F., Bomans D.J., Bothun G.D., Thompson I.B., 1995, AJ, 109, 594
\bibitem[]{}Kennicutt R.C., Lee J.C., Funes S.J., Jos\'e G., Sakai S.,  Akiyama S., 2008, ApJS, 178, 247
\bibitem[]{}Kennicutt R.C., Evans N.J., 2012, ARA\&A, 50, 531
\bibitem[]{}Koribalski B.S. et al., 2004, AJ, 1, 16
\bibitem[]{}Lauberts A., 1982, ESO/Uppsala survey of the ESO(B) atlas. Garching: European Southern Observatory
\bibitem[]{}Laustsen S., Madsen C., West R.M., 1987,  Exploring the Southern Sky.  Garching: European Southern Observatory
\bibitem[]{}Lee, J.C. et al. 2011, ApJS, 192, 33
\bibitem[]{}Li A., Draine B.T., 2001, ApJ, 554, 778
\bibitem[]{}Lindsay E.M., Shapley H., 1960, Observatory, 80, 223
\bibitem[]{}Longmore A.J. et al., 1982, MNRAS, 200, 325    
\bibitem[]{}Madore B.F., Nelson E., Petrillo K., 2009, ApJS, 181, 572   
\bibitem[]{}Marston A.P.,  Appleton P.N., 1995, AJ, 109, 1002
\bibitem[]{}Masci F., 2013, Astrophysics Source Code Library, 2010   
\bibitem[]{}Masci F.J., Fowler J.W., 2009, Astronomical Data Analysis Software and Systems XVIII, 411, 67
\bibitem[]{}McDonald I., Zijlstra A.A., Boyer M.L., 2012, MNRAS, 427, 343
\bibitem[]{}Melbourne J., Boyer M.L., 2013, ApJ, 764, 30
\bibitem[]{}Meyer M.J. et al., 2004, MNRAS, 350, 1195
\bibitem[]{}Murphy T., Mauch T., Green A., Hunstead R.W., Piestrzynska B., Kels A.P., Sztajer P., 2007, MNRAS, 382, 382
\bibitem[]{}Parker Q.A., Bland-Hawthorn J., 1998, PASA, 15, 33
\bibitem[]{}Parker Q.A. et al., 2005, MNRAS, 362, 689
\bibitem[]{}Parker Q.A. et al., 2006, MNRAS, 372, 79
\bibitem[]{}Pilyugin L.S., Vilchez J.M., Thuan, T.X., 2010, ApJ 720, 1738
\bibitem[]{}Planck Collaboration: Ade P.A.R. et al., 2014, A\&A, 571, 28   
\bibitem[]{}Rela\~no M., Kennicutt R.C., Eldridge J.J., Lee J.C., Verley S., 2012, MNRAS, 423, 2933
\bibitem[]{}Rodgers A.W., Harding P., Bloxham G., Bessell M.S., 1998, PASP, 100, 626
\bibitem[]{}Romano R., Mayya Y.D., Vorobyov E.I., 2008, AJ, 136, 1259
\bibitem[]{}Ryder S.D., Walsh W., Malin D., 1999, PASA, 16, 84
\bibitem[]{}Sabin L. et al., 2013, MNRAS, 431, 279
\bibitem[]{}Schiminovich D., Wyder T.K., Martin D., 2007, ApJS, 173, 315 
\bibitem[]{}Schlafly E.F., Finkbeiner D.P., 2011, ApJ, 737, 103
\bibitem[]{}Schroder A., Kraan-Korteweg R.C., Henning P.A., 2009, A\&A, 505, 1049
\bibitem[]{}Skrutskie M.F. et al., 2006, AJ, 131, 1163
\bibitem[]{}Sersic J.L., 1974, Astrophys. Sp. Sci., 28, 365
\bibitem[]{}Smith R.T., 1941, PASP, 53, 187
\bibitem[]{}Smith R., Lane R.R., Conn B.C., Fellhauer M., 2012, MNRAS, 423, 543
\bibitem[]{}Stark D.V., Kannappan S.J., Wei L.H., Baker A.J., Leroy A.K., Eckert K.D., Vogel S.N., 2013, ApJ, 769, 82
\bibitem[]{}Staveley-Smith L., Kim S., Calabretta M.R., Haynes R.F., Kesteven M.J., 2003, MNRAS, 339, 87
\bibitem[]{}Strauss M.A., Huchra J.P., David M., et al., 1992, ApJS, 92, 29
\bibitem[]{}Taylor E.N. et al., 2011, MNRAS, 418,1587
\bibitem[]{}Theys J.C., Spiegel E.A., 1976, ApJ, 208, 650
\bibitem[]{}Theys J.C., Spiegel E.A., 1977, ApJ, 212, 616
\bibitem[]{}Tully R.B., 1988,  Nearby Galaxies Catalog.  Cambridge and New York: Cambridge University Press
\bibitem[]{}van der Marel R., Kallivayalil N., Besla G., 2009, in Proc. IAU Symp. 256, The Magellanic System: Stars, Gas, and Galaxies, J.T. van Loon \& J.M. Oliveira, eds., p 81
\bibitem[]{}Viironen K., Delgado-Inglada G., Mampaso A., Magrini L., Corradi R.L.M., 2007, MNRAS, 381, 1719
\bibitem[]{}Vogt F.P.A., Dopita M.A., Borthakur S., Verdes-Montenegro L., Heckman T.M., Yun M.S., Chambers K.C., 2015, MNRAS in press (arXiv:1504.03337)
\bibitem[]{}Wright A.E., Griffith M.R., Burke B.F., Ekers R.D., 1994, ApJ. Suppl. Ser. 91, 111
\bibitem[]{}Woudt P.A., Kraan-Korteweg R.C., Fairall A.P., 1999, A\&A, 352, 39
\bibitem[]{}Woudt P.A, Kraan-Korteweg R.C., 2001, A\&A, 380, 441  
\bibitem[]{}Wright E.l. et al., 2010, AJ, 140, 1868
\bibitem[]{}Yamamura I. et al., 2009, ASP Conf. Ser., 418, 3
\bibitem[]{}Zwicky F., 1941, in Theodore von K\'arm\'an Anniversary Volume: Contributions to Applied Mechanics and Related Subjects.  Pasadena: California Institute of Technology. p. 137

\end{thebibliography}
\end{document}